\documentclass[12pt,english,aps,prb]{revtex4-2}
\pdfoutput = 1
\usepackage[T1]{fontenc}
\usepackage{textcomp}
\usepackage[latin9]{inputenc}
\usepackage{color}
\usepackage{float}
\usepackage{booktabs}
\usepackage{multirow}
\usepackage{amsmath}
\usepackage{amssymb}
\usepackage{stackrel}
\usepackage{graphicx}

\makeatletter

\newcommand{\lyxmathsym}[1]{\ifmmode\begingroup\def\b@ld{bold}
  \text{\ifx\math@version\b@ld\bfseries\fi#1}\endgroup\else#1\fi}

\providecommand{\tabularnewline}{\\}

\usepackage{babel}

\makeatother

\usepackage{babel}
\begin{document}
\title{Electric field induced half-metallicity in a two-dimensional ferromagnetic
Janus VSSe bilayer}
\author{Khushboo Dange}
\email{khushboodange@gmail.com}

\affiliation{Department of Physics, Indian Institute of Technology Bombay, Powai,
Mumbai 400076, India}
\author{Shivprasad S. Shastri}
\email{shastri1992@gmail.com}

\affiliation{Department of Physics, Indian Institute of Technology Bombay, Powai,
Mumbai 400076, India}
\author{Alok Shukla}
\email{shukla@iitb.ac.in}

\affiliation{Department of Physics, Indian Institute of Technology Bombay, Powai,
Mumbai 400076, India}
\begin{abstract}
Two-dimensional half-metals with intrinsic ferromagnetism hold great
potential for applications in spintronics. In this study, we aim to
expand the known space of such two-dimensional ferromagnetic (FM)
half-metals by investigating a bilayer of Janus VSSe, an FM semiconductor.
Its structural, electronic, and magnetic properties are examined using
first-principles density functional theory (DFT) based approach, employing
the DFT+\textit{U} methodology, coupled with the PBE functional. The
stability of the bilayer is examined using ab initio molecular dynamics
simulations at finite temperatures up to 400 K. To ensure the stability
further, the elastic constants of the system have also been investigated
and we found that VSSe bilayer remains stable against structural deformation.
The magnetic anisotropy calculations suggest that the VSSe bilayer
manifests an easy plane of magnetization similar to its monolayer
counterpart. At the DFT+\textit{U} level of theory, the considered
VSSe bilayer exhibits a tendency towards half-metallicity with a small
band gap of 0.11 eV for the majority spin carriers, and of 0.60 eV
for the minority ones. To induce a transition from a semiconductor
to a half-metal, the bilayer is subjected to an external electric
field of varying strengths normal to the plane. The lack of horizontal
mirror symmetry in the bilayer allows bidirectional tuning of the
band gap, with different values for the field in ``upward'' and
``downward'' directions. The band gaps for the two spin channels
increase with the increasing upward applied electric field, while
the opposite happens for the downward fields, with the majority carrier
gap closing at $\approx$ 0.16 V/{\AA}, making the material a spin
gapless semiconductor. Further increase in the electric field renders
the material half metallic at $\approx$ 0.18 V/{\AA}. Given the
fact that these values of the external electric field are achievable
in the lab suggests that the FM Janus VSSe bilayer is a promising
candidate for spintronic devices. 
\end{abstract}
\maketitle

\section{Introduction}

Spintronics is an emerging field that exploits the electron's spin
in addition to its charge for advanced device applications. For such
applications, the devices preferably require materials with 100 \%
spin polarization at the Fermi level, which is satisfied by the half-metallic
ferromagnetic materials. These properties make them ideal for developing
highly efficient spintronic components such as spin valves, magnetic
tunnel junctions, and spin transistors \cite{Ahn2020}. Spintronics
has the potential to revolutionize various technologies, including
magnetic storage, data processing, and quantum computing, with reduced
power consumption, high processing speed, and enhanced device functionality
\cite{cortie2020,Liu2019}. Two-dimensional (2D) half-metallic ferromagnets
are of great research interest in this field since they lead to miniaturization
of devices in many important applications, in addition to the full
spin polarization at the Fermi level offered by their bulk counterparts.
Despite these promising applications, intrinsic 2D ferromagnetic (FM)
half-metals are rare, with no practical realizations yet \cite{han2023,zhang2020,Yao2021,Wu2018,Ashton2017,Wang2019}.
Consequently, research into 2D magnetic materials that can be engineered
into FM half-metals in the presence of external perturbations is at
the forefront of the field of spintronics.

The transition from a semiconductor to a metal, or half-metal, can
be achieved through doping \cite{Wu2018,Pei2018}, strain engineering
\cite{Scalise2012}, or by applying an external electric field \cite{Pei2018,Rama2011,Yang_2020}.
Applying an external electric field as a perturbation is an effective
method for tuning the band gap because it provides excellent control
to the experimentalist, and does not require any changes to the material
under investigation. The reduction of the band gap by applying an
external electric field has been theoretically predicted in bilayer
structures of hexagonal boron nitride \cite{Yang2010}, and transition-metal
dichalcogenides \cite{Rama2011}, while for ZnO thin films it has
been experimentally demonstrated \cite{imt-zno-2018}. Recent studies
reported that 2D fully compensated ferrimagnetic materials with zero
net magnetization can also show fully spin-polarized currents in the
half-metallic states upon the application of an external electric
field \cite{San-Dong1,San-Dong2}. The transition from a semiconducting
phase to a half-metal has been reported for the FM CrSBr monolayer
and CrI$_{3}$-CrGeTe$_{3}$ heterobilayer by applying external electric
fields of strengths close to 0.4 V/{\AA} \cite{Guo2023}, and 0.6
V/{\AA} \cite{Tang2020}, respectively. However, the practical
realization of such high electric field strengths is normally challenging,
which drives the search for 2D magnetic materials where half-metallic
behavior can be achieved at lower electric field strengths.

The family of 2D transition-metal dichalcogenides (TMDs) represents
an emerging class of versatile materials, exhibiting a wide range
of interesting electronic, optical, and magnetic properties \cite{Duan2015}.
First-principles calculations have predicted intrinsic long-range
FM order in TMDs, when the underlying transition metal is V, Fe, or
Mn \cite{zhuang2016,Ma2012,chen2020}. The practical realization of
long-range FM order in 2D magnets is challenging according to the
Mermin-Wagner theorem, which states that magnetic ordering gets suppressed
over long ranges when reduced to lower dimensions at any finite temperature
\cite{mermin1996}. Despite this, several atomically thin 2D van der
Waal (vdW) magnets have been successfully synthesized, exhibiting
long-range FM order \cite{Huang2017,Gong2017,OHara2018}. This is
due to the strong magnetic anisotropy in 2D magnetic materials, which
provides thermal stability thus sustaining long-range magnetic order.
Bonilla \textit{et al.} \cite{Bonilla2018} provided the first experimental
evidence of intrinsic long-range FM order in a VSe$_{2}$ monolayer,
marking it as the first 2D TMD to exhibit room-temperature ferromagnetism.
The presence of this long range FM order makes the VSe$_{2}$ monolayer
a promising candidate for spintronic devices as also suggested by
theoretical studies \cite{Ma2012}. Similarly, based on first-principles
calculations, the VS$_{2}$ monolayer has been proposed for spintronic
applications due to its predicted long-range FM order \cite{zhuang2016,Ma2012}.

Janus TMDs which are a new class of materials with broken horizontal
mirror symmetry due to different chalcogen atoms on each side exhibit
spontaneous polarization. This facilitates enhanced band gap tunability
and Rashba spin splitting both of which are beneficial for spintronic
and other applications \cite{Hu2018,Riis-Jensen2019,Zhang2019}. The
successful synthesis of 2D non-magnetic Janus MoSSe on the SiO$_{2}$/Si
substrate \cite{Zhang2017} has sparked an upsurge in research interest
in the realm of 2D Janus materials, including vanadium-based Janus
monolayers \cite{junjie2018,Zhang2019,Dey2020}. Janus VSSe monolayer
is one such semiconductor, predicted to be stable in both the 2H and
1T phases, with the 2H phase and corresponding FM state being energetically
more stable \cite{Zhang2019,Dey2020}. Its dynamical stability has
been reported in the literature by the absence of imaginary phonon
frequencies \cite{Zhang2019,Dey2020}, while ab initio molecular dynamics
calculations have shown its thermal stability up to 500 K \cite{Luo2020}.
Due to broken time-reversal, inversion, and mirror symmetries, the
FM Janus VSSe monolayer has strong response to external stimuli \cite{Zhang2019},
thereby allowing easy tunability. As FM half-metals are superior for
application in spintronics, FM semiconductor to FM half-metal transition
of the 2D Janus VSSe will be particularly advantageous. This transition
can be induced by an external electric field if the structure consists
of a bilayer of magnetic atoms and the energy bands near the Fermi
level correspond to the same spin channel, with different layer characteristics
\cite{Guo2023}. Therefore, in this work, from the perspective of
applications in spintronics, we explore a bilayer of 2D Janus VSSe
composed of different interfacial chalcogen layers, thereby retaining
the broken symmetries of the monolayer. The properties of 2D Janus
VSSe bilayer have not been studied in detail in the literature. For
the purpose of benchmarking, we initiate our investigation by computing
some intrinsic properties of the FM Janus VSSe monolayer, and compare
them with the existing literature. Next, we simulate the bilayer formation
by exploring three different stacking arrangements of the monolayers,
and find that the most stable arrangement also exhibits FM order.
Following this, we studied its structural, mechanical, electronic,
and magnetic properties using first-principles density functional
theory. The obtained band gap is lower compared to that of the monolayer
for both the majority and minority spin carriers. We have explored
the bidirectional tuning of the band gap by changing the magnitude
and the direction of the perpendicular external electric field. We
find that the external field in the ``upward'' direction increases
the band gap, while, reversing the field has the opposite effect.
In the latter case, at the applied field of $\approx$ 0.18 V/{\AA}
, the gap closes for the majority spin carriers and a significant
gap persists for the minority ones, resulting in a half-metal.

The remainder of this paper is organized as follows. The next section
deals with the discussion of the employed computational methods. Section
\ref{sec:results-and-discussion} is dedicated to the presentation
and discussion of our findings which are summarized and concluded
in the last section.

\section{Computational Methodology}

Given the fact that the considered material is magnetic in nature,
all the calculations were spin-polarized, and performed using the
Vienna Ab initio Simulation Package (VASP) \cite{kresse1996,Georg1996}
within the framework of the state-of-the-art density functional theory
(DFT) \cite{hohenberg1964,kohn1965}. The kinetic energy cutoff has
been set to 600 eV for the expansion of the Bloch wave functions,
using a plane wave basis set. The valence electronic configurations
of V, S, and Se atoms are 3$d^{3}$4$s^{2}$, 3$s^{2}$3$p^{4}$,
and 4$s^{2}$4$p^{4}$, respectively, within the projector augmented
wave (PAW) approach \cite{kresse1999,blochl1994}. To account for
the vdW interactions in the case of bilayer VSSe, Grimme's DFT-D3
\cite{grimme2010} dispersion correction has been adopted. To avoid
the interaction of the considered 2D VSSe structures with their periodic
replicas, the length of \textit{c} lattice parameter was set to 20
{\AA} and 25 {\AA} for the monolayer and bilayer, respectively.
The convergence criterion for geometry optimization has been set to
0.01 eV/{\AA} for the Hellmann-Feynman forces. The energy convergence
threshold of $10^{-5}$ eV has been set for the self-consistent-field
iterations in solving the Kohn-Sham equations. The Monkhorst-Pack
scheme \cite{monkhorst1976} has been employed to perform the Brillouin
zone integration, for which, the \textbf{k}-mesh of the size $12\times12\times1$
is employed for the geometry relaxation of the considered Janus VSSe
monolayer and bilayers. Afterwards, the \textbf{k}-mesh is refined
to $30\times30\times1$ for further calculations on the optimized
monolayer and bilayers. The generalized gradient approximation (GGA)
has been considered for the exchange-correlation functional. We start
by using the Perdew-Burke-Ernzerhof (PBE) \cite{perdew1996} version
of the GGA for all the optimizations and to determine the structural
properties of both the Janus VSSe monolayer and bilayers. Then, to
better address the magnetic and electronic properties, we use the
rotationally invariant DFT+\textit{U} (GGA+\textit{U}) approach introduced
by Liechtenstein \textit{et al.} \cite{Liechtenstein1995}. GGA+\textit{U}
is known to yield improved results as compared to GGA for the systems
involving correlated electrons by taking into account the on-site
Coulomb repulsion experienced by the localized electrons of \textit{d}
orbitals. The effective on-site interaction has been added to the
3$d$ electrons of vanadium by taking the Hubbard $U$ parameter equal
to 2.7 eV. The anisotropy of exchange interactions has been incorporated
by taking Hund's coupling $J=0.7$ eV. Our considered $U$ and $J$
values are consistent with the values used to study vanadium-based
systems (non-Janus and Janus dichalcogenides) in earlier studies \cite{Li2014,zhuang2016,Dey2020,aryasetiawan2006}.
The Bader charge analysis has been performed using a code developed
by Henkelman group \cite{tang2009} to quantify the charges present
on the involved atomic species.

The thermal stability of the VSSe bilayer was investigated by performing
ab initio molecular dynamics (AIMD) simulations \cite{AIMD1994}.
These simulations were conducted in the canonical ensemble using a
Nos�-Hoover thermostat at two temperatures: room temperature (300
K) and an elevated temperature of 400 K. To determine the mechanical
stability of the VSSe bilayer, the equilibrium lattice configuration
has been subjected to small loading strains ranging from -2 \% to
2\% in increments of 0.5 \%. We focused on elastic deformations within
the harmonic range in which the material exhibits a linear stress
($\sigma$) response to the applied strain ($\epsilon$), governed
by the Hooke's law. This relation is expressed in the Voigt notation
\cite{Andrew2012} and is given by 
\begin{equation}
\sigma_{i}=\sum_{j=1}^{6}C_{ij}\epsilon_{j},\label{eq:hookeslaw}
\end{equation}

where $\sigma_{i}$ and $\epsilon_{j}$ represent the six independent
components of the stress and strain vectors, respectively (1 $\leq$
i, j $\leq$ 6), and $C_{ij}$ is the second-order elastic stiffness
tensor expressed as a 6 $\times$ 6 symmetric matrix. The elastic
energy $\Delta E(V,\{\epsilon\})$ of a strained structure under the
harmonic approximation, is defined as

\begin{equation}
\Delta E(V,\{\epsilon_{i}\})=E(V,\{\epsilon_{i}\})-E(V_{0},0)=\frac{V_{0}}{2}\sum_{i,j=1}^{6}C_{ij}\epsilon_{i}\epsilon_{j},\label{eq:energy-strain}
\end{equation}

where $E(V,\{\epsilon_{i}\})$ and $E(V_{0},0)$ denote the total
energies of the strained and unstrained crystal structures with volumes
$V$ and $V_{0}$, respectively. For 2D materials in the $xy$-plane,
the stiffness tensor of Eq. \ref{eq:hookeslaw} reduces to a 3 \texttimes{}
3 symmetric matrix, which gives in-plane stiffness components $C_{ij}$
with i, j = 1, 2, 6. These second order $C_{ij}$ components are determined
using the energy-strain relation as implemented in the VASPKIT package
\cite{WANG2021}. For a 2D hexagonal isotropic system in the $xy$-plane,
the relations simplify to $C_{61}=C_{16}=C_{62}=C_{26}=0$, leaving
only two independent constants, $C_{11}$ and $C_{12}$ where $C_{11}=C_{22}$,
$C_{21}=C_{12}$, and $C_{66}=0.5\times(C_{11}-C_{12})$. Additionally,
the direction-dependent Young's modulus $E(\theta)$ and the Poisson's
ratio $\nu(\theta)$ were calculated using the following equations:
\begin{equation}
E(\theta)=\frac{C_{11}C_{22}-C_{12}^{2}}{C_{11}\sin^{4}\theta+\left(\frac{C_{11}C_{22}-C_{12}^{2}}{C_{66}}-2C_{12}\right)\sin^{2}\theta\cos^{2}\theta+C_{22}\cos^{4}\theta},\label{eq:youngs}
\end{equation}

and 
\begin{equation}
\nu(\theta)=\frac{C_{12}(sin^{4}\theta+cos^{4}\theta)-(C_{11}+C_{22}-\frac{C_{11}C_{22}-C_{12}^{2}}{C_{66}})\sin^{2}\theta\cos^{2}\theta}{C_{11}\sin^{4}\theta+\left(\frac{C_{11}C_{22}-C_{12}^{2}}{C_{66}}-2C_{12}\right)\sin^{2}\theta\cos^{2}\theta+C_{22}\cos^{4}\theta}\label{eq:poisson}
\end{equation}

\section{results and discussion}

\label{sec:results-and-discussion}

We started the calculations by optimizing the geometry of the VSSe
monolayer. Next, we calculated the spin-polarized electronic band
structure, density of states (DOS), and the magnetic moments for the
FM monolayer using the GGA+\textit{U} approach, and found that our
results are consistent with the previous studies \cite{Dey2020}.
More details about the crystal structure, electronic band structure,
DOS, magnetic moments, and formation energy for the monolayer are
provided in the Supplemental Material (SM) \cite{supporting-data},
and some references cited therein \cite{Qi2020,Wasey2015,wang2018}.
Next, we present and discuss results of our calculations on the bilayer.

\subsection{Stacking configurations and stability}

\begin{figure}[H]
\begin{centering}
\includegraphics[scale=0.35]{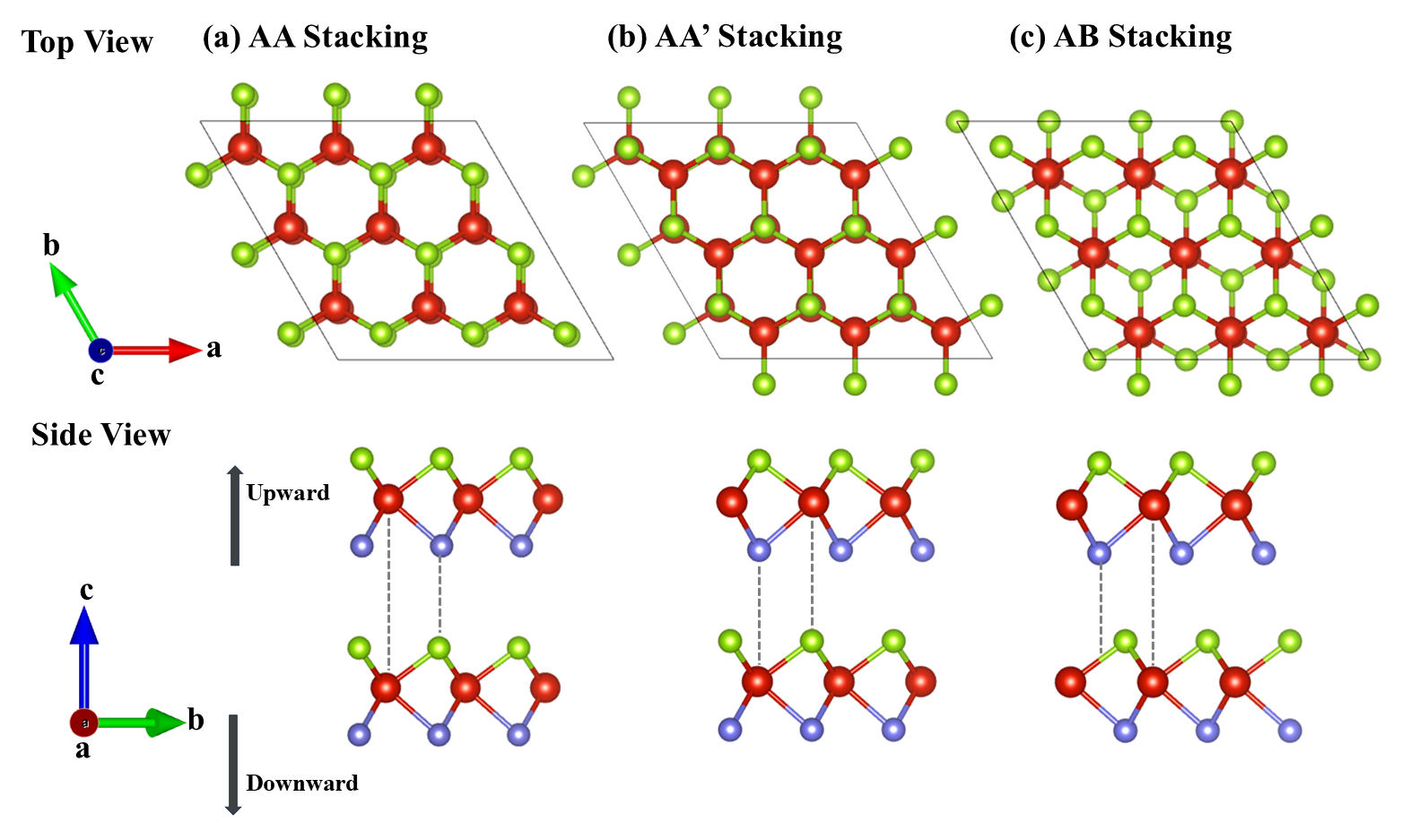} 
\par\end{centering}
\caption{Considered stacking configurations of the bilayer of Janus VSSe. Red,
green, and blue spheres represent the V, S, and Se atoms, respectively. }
\label{fig:bilayer-stack} 
\end{figure}

In order to form the bilayer of VSSe, we stacked two geometry-optimized
monolayers on top of each other in the following orders: (a) AA stacking
(Fig. \ref{fig:bilayer-stack}(a)) with the same type of atoms on
both top and bottom layers, (b) AA' stacking (Fig. \ref{fig:bilayer-stack}(b))
in which the top layer is anti-parallel to the bottom layer with V
atoms vertically above the chalcogen atoms, S and Se of the bottom
layer (and vice versa), and (c) finally AB stacking shown in Fig.
\ref{fig:bilayer-stack}(c) which is obtained by shifting the layers
of AA' stacking. The AA stacked VSSe bilayer with broken inversion
symmetry belongs to \textit{P3m1} space group and possesses the $C_{3v}$
point group. In AA' stacking, a layer is rotated by 180$^{\circ}$
relative to another layer, which gives rise to a centrosymmetric bilayer
if the two chalcogens are the same. However, in the case of VSSe,
the two different chalcogens lead to the non-centrosymmetric AA' stacking.
The AB stacking, which is also non-centrosymmetric, is obtained by
shifting the layers of AA' stacking along the armchair direction such
that V atoms of two layers are vertically above each other and the
chalcogen of one layer lies on the top of the hexagon center of another
layer. Due to the lack of inversion symmetry, both the AA' and AB
stackings also belong to \textit{P3m1} space group, similar to the
AA stacked VSSe bilayer. Thus, all three considered VSSe bilayers
belong to the same space group as the monolayer. The AA' stacking
type is normally observed in the bilayer of 2H-TMDs such as bilayer
2H-MoS$_{2}$ \cite{He2014}. The monolayer of Janus 2H-VSSe has FM
ordering in its ground state. When a bilayer composed of two such
monolayers is formed, different magnetic orderings can be achieved
with the change in the type of stacking leading to different ground
states. For this purpose, we used 2$\times$2 supercells consisting
of 24 atoms, and built an FM and six possible antiferromagnetic (AFM)
configurations considering various intra- and interlayer FM and AFM
arrangements among V atoms. Further, supercells of various configurations
are relaxed to obtain the optimized geometries including the interlayer
distance corresponding to the ground state. The optimized FM and AFM
configurations (AFM1 to AFM6) are shown in Fig. S5 of SM \cite{supporting-data},
where a discussion on the nature of magnetic order of each configuration
is also presented.

After optimization we found that FM ordering is favored as ground
state in all three stacking configurations. The energy difference
with respect to corresponding ground state, $\Delta E$, and interlayer
distance, $d$ (distance between two metal atoms of the top and bottom
layers) after optimization for different stacking and magnetic configurations
are given in Table \ref{tab:bilayer-geometry}. The energy difference
is calculated as $\Delta E=E_{AFM}-E_{FM}$, where, $E_{FM}$ and
$E_{AFM}$ represent the total energies of the FM and different AFM
configurations, respectively. The small energy difference ($\sim$
5 meV) between FM and AFM1 configurations is obtained for each stacking
because the two configurations differ only in their interlayer magnetic
interactions which are weak, and the strong intralayer interactions
in both cases are same (ferromagnetic). In contrast, other configurations,
AFM2 -- AFM6 involve antiferromagnetic interactions both interlayer
and intralayer, leading to significantly larger energy differences
relative to the FM configuration. The total energies of the three
considered stacking configurations in the FM state follow the order:
$E_{AB}<E_{AA'}<E_{AA}$. Among them, the AB stacking exhibits the
lowest energy, indicating it is the most stable configuration. Based
on this energy trend, the relative stability of the stackings with
FM ordering is: AB > AA' > AA. The trend in interlayer distance $d$
is $d_{AA}>d_{AA'}>d_{AB}$, with the minimum $d$ of 6.05 {\AA}
for the AB stacking. The interlayer distance shows the same trend
as in the order of energies of different stackings for FM ordering.
Following this, we proceed with a detailed investigation of the FM
configuration of the AB stacked bilayer including the band gap tunability
with respect to an external electric field.

\begin{table}[H]
\caption{Interlayer distance, $d,$ and the relative energy, $\Delta E$ of
various magnetic configurations for different stackings. }
\label{tab:bilayer-geometry}

\centering{}%
\begin{tabular}{ccccccccc}
\toprule 
\multirow{1}{*}{Stacking} & \multirow{1}{*}{$d$ ({\AA})} & \multicolumn{7}{c}{$\Delta E$ (meV per unit cell)}\tabularnewline
\midrule 
 &  & FM  & AFM1  & AFM2  & AFM3  & AFM4  & AFM5  & AFM6\tabularnewline
\midrule
\midrule 
AA  & 6.67  & 0  & 2.10  & 94.65  & 94.15  & 95.57  & 94.22  & 96.51\tabularnewline
AA'  & 6.20  & 0  & 3.38  & 100.40  & 100.00  & 101.83  & 100.45  & 104.36\tabularnewline
AB  & 6.05  & 0  & 4.59  & 94.39  & 92.61  & 92.89  & 92.03  & 93.33\tabularnewline
\bottomrule
\end{tabular}
\end{table}

\begin{figure}[H]
\begin{centering}
\includegraphics[scale=0.3]{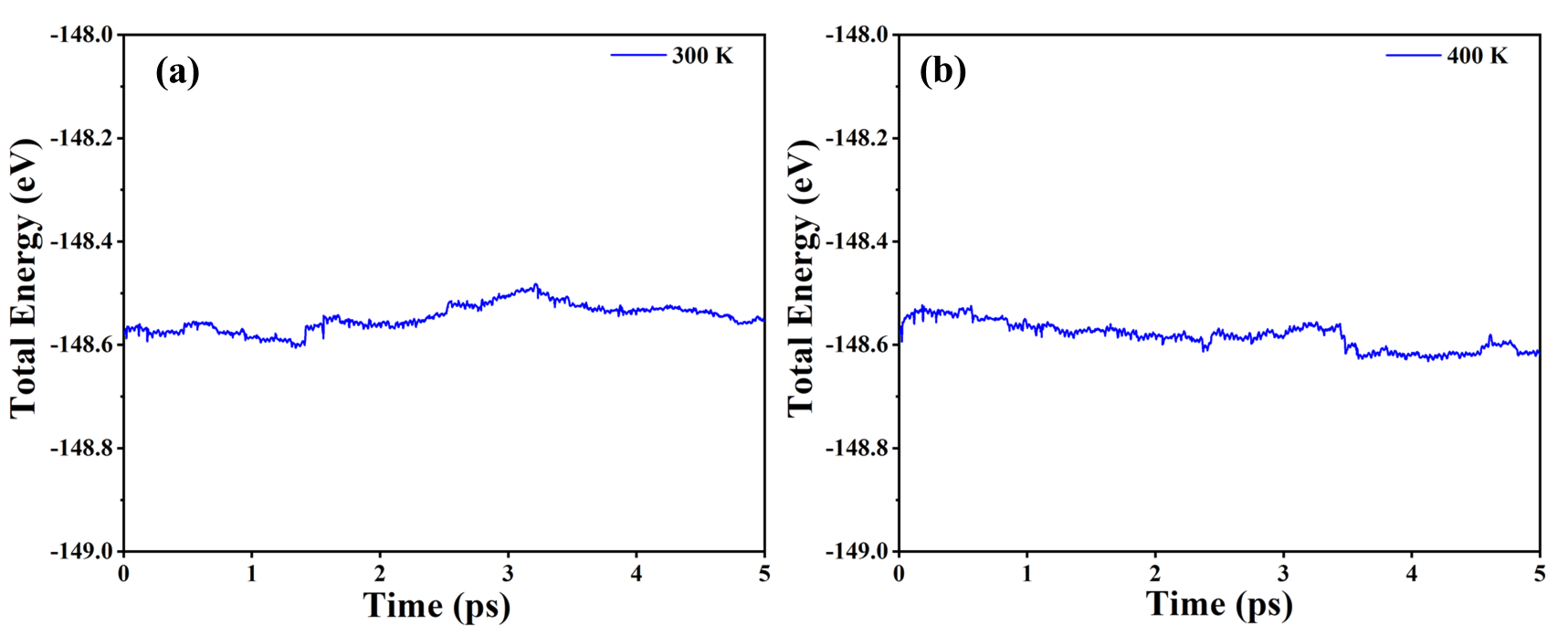} 
\par\end{centering}
\caption{Total energy of the considered VSSe bilayer as a function of time
at (a) 300 K and (b) 400 K. }
\label{fig:md_plots} 
\end{figure}

We calculated the layer binding energy, $E_{b}$ as 
\begin{equation}
E_{b}=E_{VSSe}^{AB}-2E_{VSSe}
\end{equation}

where $E_{VSSe}$ and $E_{VSSe}^{AB}$ represent the total energies
of the considered FM VSSe monolayer and bilayer, respectively. The
obtained $E_{b}$ of -1.09 eV suggests the feasibility of bilayer
formation in the AB stacking from the constituent monolayers. Further,
to assess the thermal stability of the VSSe bilayer, we performed
AIMD simulations in the canonical ensemble, employing the Nos�-Hoover
thermostat \cite{Nose1984}. The AIMD simulations were carried out
separately at 300 K and 400 K, each for a total duration of 5 ps with
a time step of 1 fs. Fig. \ref{fig:md_plots} depicts the variation
of the total energy over time for both the temperatures, and it is
observed that there are minimal fluctuations in total energy (less
than 0.1 eV) and minor changes in bond lengths, with the maximum variations
of 0.11 {\AA} and 0.10 {\AA} for V-S and V-Se bonds, respectively,
indicating the stability of the Janus VSSe bilayer at both the room
temperature and elevated temperatures.

\subsection{Mechanical properties }

\begin{figure}[H]
\begin{centering}
\includegraphics[scale=0.4]{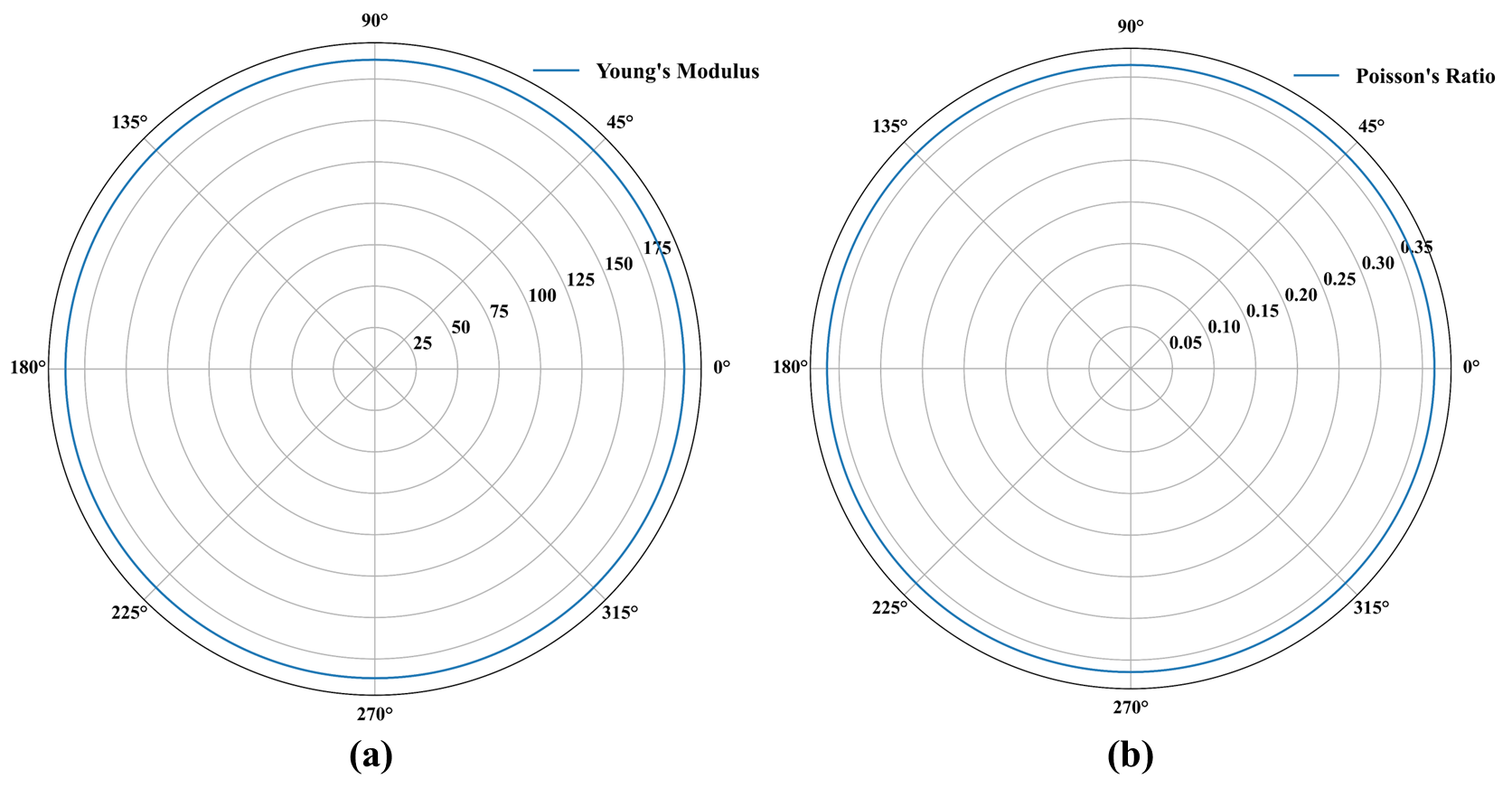} 
\par\end{centering}
\caption{Orientation dependent (a) Young's Modulus $E(\theta)$ in the units
of Nm\protect\protect\protect\protect\textsuperscript{-1} and
(b) Poisson ratio $\nu(\theta)$ computed for the Janus VSSe bilayer.}
\label{fig:polar-plots} 
\end{figure}

To evaluate the mechanical stability of the Janus VSSe bilayer, its
mechanical properties were systematically investigated. Mechanical
stability against structural deformation is critical, as lattice mismatch
with a substrate can induce strain. Moreover, understanding mechanical
properties is essential to assess the material's stiffness under external
strain, as strain engineering serves as an effective method for tuning
the properties of nanomaterials. To investigate these properties,
the VSSe bilayer was subjected to small deformations with strains
ranging from -2\% to 2\%, in increments of 0.5\%. The total energy
was calculated for each strain configuration, and the elastic stiffness
constants were extracted by fitting the energy-strain curves using
Eq. \ref{eq:energy-strain}. The obtained elastic stiffness constants
$C_{11}$= $C_{22}$= 215.21 Nm\textsuperscript{-1}, $C_{21}$= $C_{12}$=
78.43 Nm\textsuperscript{-1}, and $C_{66}$= 68.39 Nm\textsuperscript{-1}
satisfy the Born-Huang criteria \cite{Born1955}, $C_{11}C_{22}-C_{12}^{2}$
> 0 and $C_{66}$> 0, confirming the mechanical stability of the Janus
VSSe bilayer. Using the computed stiffness constants, additional mechanical
properties were derived such as the shear modulus $G$, direction-dependent
Young's modulus $E(\theta)$, and Poisson's ratio $\nu(\theta)$.
The shear modulus $G$ is generally equal to $C_{66}$, was determined
to be 68.39 Nm\textsuperscript{-1}, indicating significant resistance
of the VSSe bilayer to shear deformation. Further, $E(\theta)$ and
$\nu(\theta)$ are determined using Eq. \ref{eq:youngs} and Eq. \ref{eq:poisson},
with their respective polar plots depicted in Fig. \ref{fig:polar-plots}(a)
and Fig. \ref{fig:polar-plots}(b). The computed results show isotropy
in the mechanical behavior along parallel and perpendicular directions,
with $E(0^{\circ})=E(90^{\circ})$ = 186.62 Nm\textsuperscript{-1}
and $\nu(0^{\circ})=\nu(90^{\circ})$ = 0.36. This mechanical isotropy
is further reflected in the uniform Young's modulus and Poisson's
ratio across all directions in the $xy$-plane, consistent with the
bilayer's isotropic crystal structure. The calculated Young's modulus
for the Janus VSSe bilayer is lower than those reported for MoS\textsubscript{2},
MoSe\textsubscript{2}, WS\textsubscript{2}, WSe\textsubscript{2},
and WTe\textsubscript{2} bilayers but slightly higher than that of
MoTe\textsubscript{2} bilayer \cite{zeng2015}. It also exceeds the
values reported for non-magnetic Janus TMD monolayers (MXY, where
M = Mo, W and X/Y = S, Se, and Te) \cite{THANH2020}. For comparison,
the theoretically calculated Young's modulus of the recently synthesized
Janus MoSSe monolayer is 123.43 Nm\textsuperscript{-1} \cite{THANH2020}.
The positive Poisson ratio indicates that under tensile (compressive)
strain in a direction, the bilayer expands (contracts) in the perpendicular
direction, and its small value of $\nu$ = 0.36 suggests slight deformation.
Overall, the analysis confirms the mechanical stability and isotropy
of the Janus VSSe bilayer, making it a promising candidate for strain-engineered
applications.

The theoretical calculations of energetic, thermal, and mechanical
stability support the feasibility of the synthesis of the Janus VSSe
bilayer. Experimental realization of Janus MoSSe monolayer from its
non-Janus counterparts, MoS$_{2}$and MoSe$_{2}$ monolayers by thermal
selenization and sulphurization, respectively, have been reported
in the literature \cite{Zhang2017,Lu2017}. The monolayers of vanadium-based
TMDs VSe$_{2}$ and VS$_{2}$ have already been synthesized experimentally
\cite{Bonilla2018,Chen2018,vanEfferen2024}. Thus, the likelihood
of synthesizing a Janus VSSe monolayer using similar experimental
techniques appears promising. Subsequently, starting from the VSSe
monolayers, corresponding bilayer can be obtained using the transfer
techniques.

\subsection{Electronic structure and magnetic properties}

\label{subsec:Electronic-and-magnetic}

\subsubsection{Electronic Structure}

\begin{figure}[H]
\begin{centering}
\includegraphics[scale=0.4]{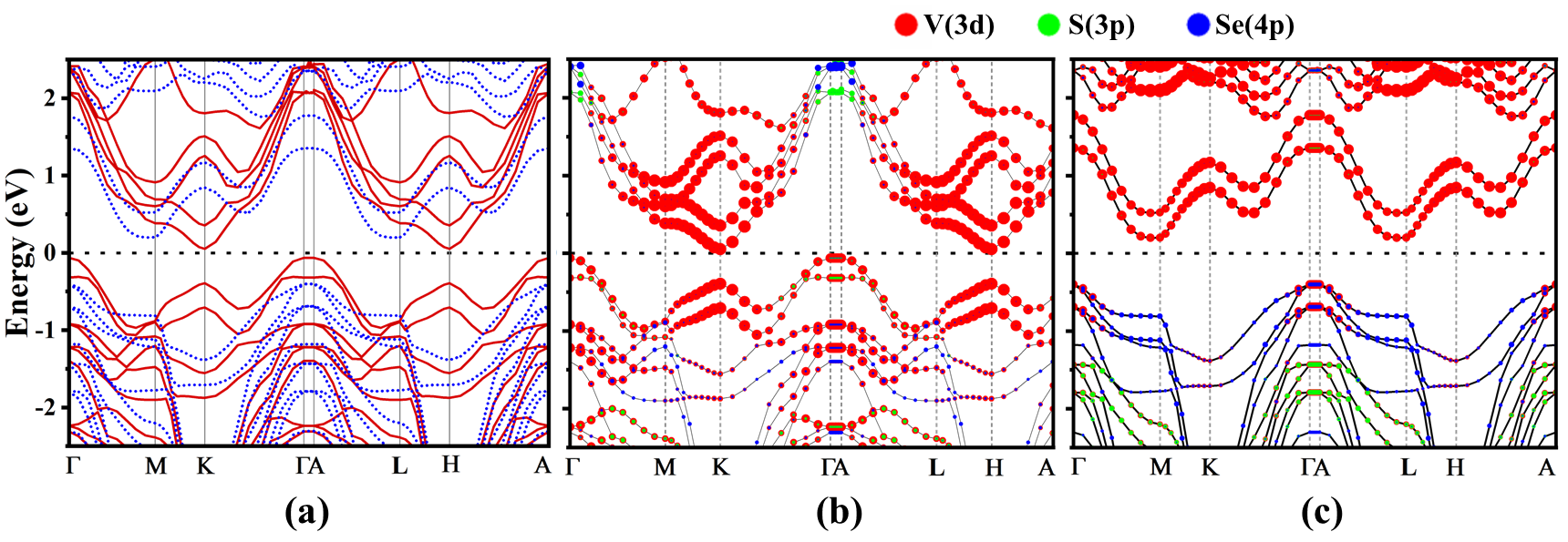} 
\par\end{centering}
\caption{ (a) Spin-polarized band structure, and projected band structures
for (b) majority and (c) minority spin channels, of the VSSe bilayer.
The solid red and dotted blue lines in (a) denote the energy states
corresponding to the majority and minority spin channels, respectively.
The dashed line at zero energy represents the Fermi energy, $E_{F}$.}
\label{fig:bilayer-band} 
\end{figure}

\begin{figure}[H]
\begin{centering}
\includegraphics[scale=0.5]{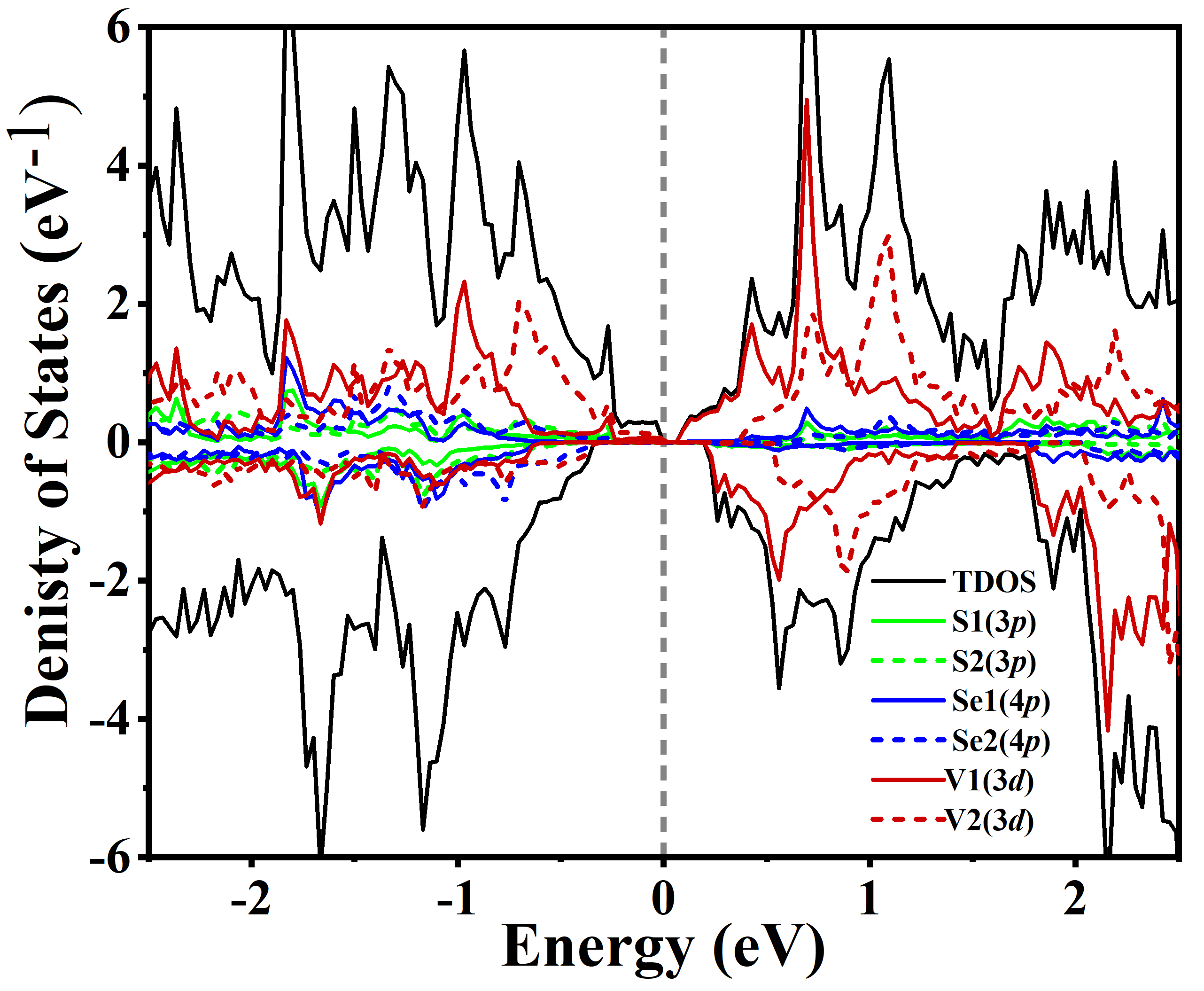} 
\par\end{centering}
\caption{ The spin-polarized TDOS and PDOS plotted for the considered FM Janus
VSSe bilayer. 1 and 2 stand for the atomic species belonging to the
bottom and top layers of the bilayer. The dashed line at zero energy
represents $E_{F}$.}
\label{element-dos} 
\end{figure}

Now we present and discuss the electronic properties of the FM VSSe
bilayer, calculated using the GGA+\textit{U} method. From the computed
spin-polarized band structure shown in Fig. \ref{fig:bilayer-band}(a)
we note that, as compared to the monolayer, the magnitude of band
gap ($E_{g}$) for the bilayer gets significantly reduced to 0.11
eV and 0.60 eV for the majority and minority spin carriers, respectively.
However, the positions corresponding to valence band maximum (VBM)
and conduction band minimum (CBM) remain unaltered for both the spin
orientations even after the formation of the bilayer, thus preserving
the indirect nature of $E_{g}$. Interestingly, the energy states
close to the Fermi level ($E_{F}$) in both the valence and conduction
band regions correspond to the same spin channel, i.e., the majority
spin channel. This feature suggests a high probability of gap closing
for majority spin bands compared to minority spin bands under the
influence of an external perturbation, rendering the bilayer VSSe
half-metallic. The 3D plots of the highest valence band and lowest
conduction band corresponding to the majority spin carriers at k$_{z}$
= 0 plane are shown in Fig. S6 of the SM \cite{supporting-data}.
In order to see the contribution of atomic orbitals to the band structure,
the atom-projected band structures corresponding to the outermost
orbitals are presented in Figs. \ref{fig:bilayer-band} (b) and (c),
for the majority and minority spin orientations, respectively. It
is observed that the bilayer resembles the monolayer VSSe in terms
of the contribution of the involved atomic species to the CBM and
VBM, corresponding to both the majority and minority spin carriers
(See Fig. S3 of the SM \cite{supporting-data} for atom-projected
band structures of the monolayer). Furthermore, we show the spin-polarized
atom-projected partial density of states (PDOS) in Fig. \ref{element-dos}.
The atoms of the bottom layer are labeled as V1, S1, and Se1, while
those of the top layer are labeled as V2, S2, and Se2. It can be observed
from Fig. \ref{element-dos} that VBM has dominant top layer character
(V2 (3\textit{d})) and CBM shows bottom layer characteristics (V1
(3\textit{d})) for the majority spin carriers. The CBM and VBM of
minority spin carriers also possess the bottom and top layer characteristics,
respectively, however, the Se2 (4\textit{p}) atom also contributes
to VBM along with the V2 (3\textit{d}) atom. The contribution from
different layers to the CBM and VBM of the spin channel close to the
$E_{F}$ is important for achieving FM semiconductor-to-half metal
transition in the presence of external stimuli, as stated earlier.
The majority spin channel of the considered VSSe bilayer satisfies
this criterion as the contribution to CBM comes from the V(3\textit{d})
atom of one layer, while the contribution to the VBM comes from the
V (3\textit{d}) atom of another layer, thereby resulting in different
layer characteristics. The total density of states (TDOS) plotted
in the same figure (Fig. \ref{element-dos}) is consistent with the
calculated band structure of the VSSe bilayer. 

\begin{figure}
\begin{centering}
\includegraphics[scale=0.4]{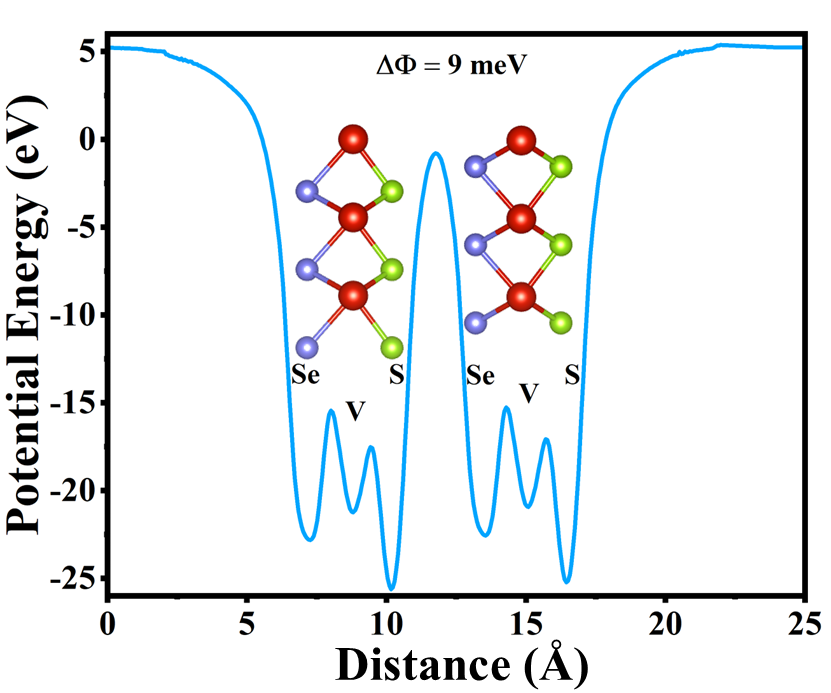} 
\par\end{centering}
\caption{Planar average of the electrostatic potential energy as a function
of distance in the z-direction, plotted for the VSSe bilayer under
consideration. $\Delta\Phi$ represents the change in work function
and its non-zero value confirms the presence of inherent $\mathbf{E_{in}}$. }
\label{fig:bi-potential} 
\end{figure}

Similar to the VSSe monolayer, the structure of the VSSe bilayer shows
a lack of horizontal mirror symmetry, and a difference in electronegativity
between chalcogen atoms which can give rise to spontaneous polarization,
\textbf{P}, and an internal electric field $\mathbf{E_{in}}$. To
illustrate the presence of the intrinsic electric field and polarization,
we plotted the planar average of electrostatic potential energy perpendicular
to the 2D surface, as a function of distance along $z$-direction
(see Fig. \ref{fig:bi-potential}). The asymmetric nature of the electrostatic
potential energy for each layer indicates the presence of a non-vanishing
$\mathbf{E_{in}}$. Further, the change in work function, $\Delta\Phi=9$
meV is also observed as we move from the lower layer to the upper
one, consistent with the presence of $\mathbf{E_{in}}$ and spontaneous
polarization in the bilayer VSSe. Note that the value of $\Delta\Phi$
in the case of the bilayer is more than that obtained for the monolayer
case (5 meV). It is worth noting that the magnitude of \textbf{P}
increases with increasing $\Delta\Phi$. Therefore, the value of \textbf{P}
should be higher for the bilayer compared to the monolayer, and, indeed
\textbf{P} computed as dipole moment per unit area, is 0.04 and 0.05
\textit{e}/nm, for the mono- and bilayer, respectively. Furthermore,
\textbf{P}\textsubscript{bilayer}\textbf{ }> \textbf{P}\textsubscript{monolayer}
is consistent with the fact that the bilayer has stronger $\mathbf{E_{in}}$
as compared to the monolayer.

\begin{figure}[H]
\begin{centering}
\includegraphics[width=7cm,totalheight=10cm]{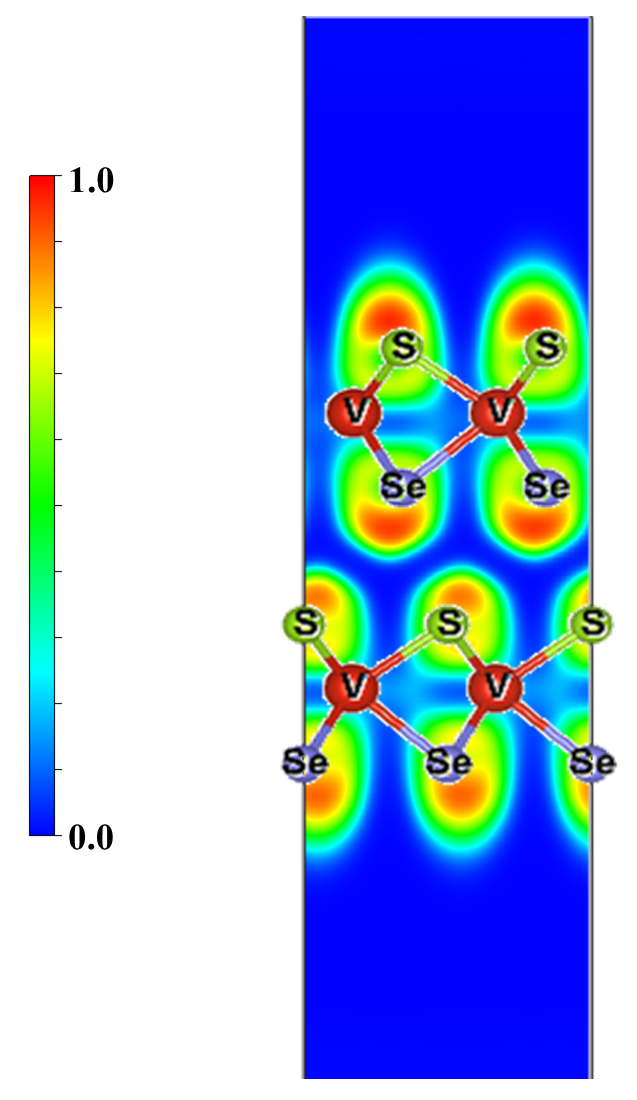} 
\par\end{centering}
\caption{Electron localization function plot for the considered VSSe bilayer
which shows the electronic distribution around atoms. The values 0.0
and 1.0 denote the fully delocalized and localized electronic charge
distribution, respectively. }
\label{fig:ELF} 
\end{figure}

The Bader charge analysis has also been performed to see the amount
of charge gain/loss by each atom in the bilayer, which is given in
Table \ref{tab:bader}. The negative and positive signs denote the
gain and the loss of electrons, respectively. The table suggests that
V atoms lose electrons, while, the chalcogen atoms gain electrons.
The amount of electronic charge gained by each S atom is more compared
to the Se atom due to the higher electronegativity of the former atom.
This leads to a redistribution of the electronic charges and creation
of an $\mathbf{E_{in}}$ in the upward direction which is also in
agreement with Fig. \ref{fig:bi-potential}. The nature of bonding
is also verified by computing the electron localization function (ELF)
presented in Fig. \ref{fig:ELF}, which takes the values between 0.0
and 1.0 denoting complete delocalization, and perfect localization
of the electrons, respectively. A highly localized electronic charge
distribution is found around the chalcogen atoms, with the delocalized
electron density around the V atom. This is attributed to the higher
electronegativity of the S/Se atoms compared to the V atom, thereby
resulting in more ionic character of the V--S and V--Se bonds with
partial covalent character as indicated by the green region (value
0.5) around them.

\begin{table}[H]
\caption{Charge gain/loss of each atomic species obtained using the Bader charge
analysis in terms of charge of an electron ($e$). Here positive and
negative signs signify the electron loss and gain, respectively.}
\label{tab:bader} \centering{}%
\begin{tabular}{cccccc}
\toprule 
\multicolumn{3}{c}{Layer 1} & \multicolumn{3}{c}{Layer 2}\tabularnewline
\midrule
\midrule 
V1  & S1  & Se1  & V2  & S2  & Se2\tabularnewline
1.25  & -0.70  & -0.55  & 1.24  & -0.70  & -0.54\tabularnewline
\bottomrule
\end{tabular}
\end{table}

\subsubsection{Magnetic Properties}

\begin{figure}[H]
\begin{centering}
\includegraphics[scale=0.4]{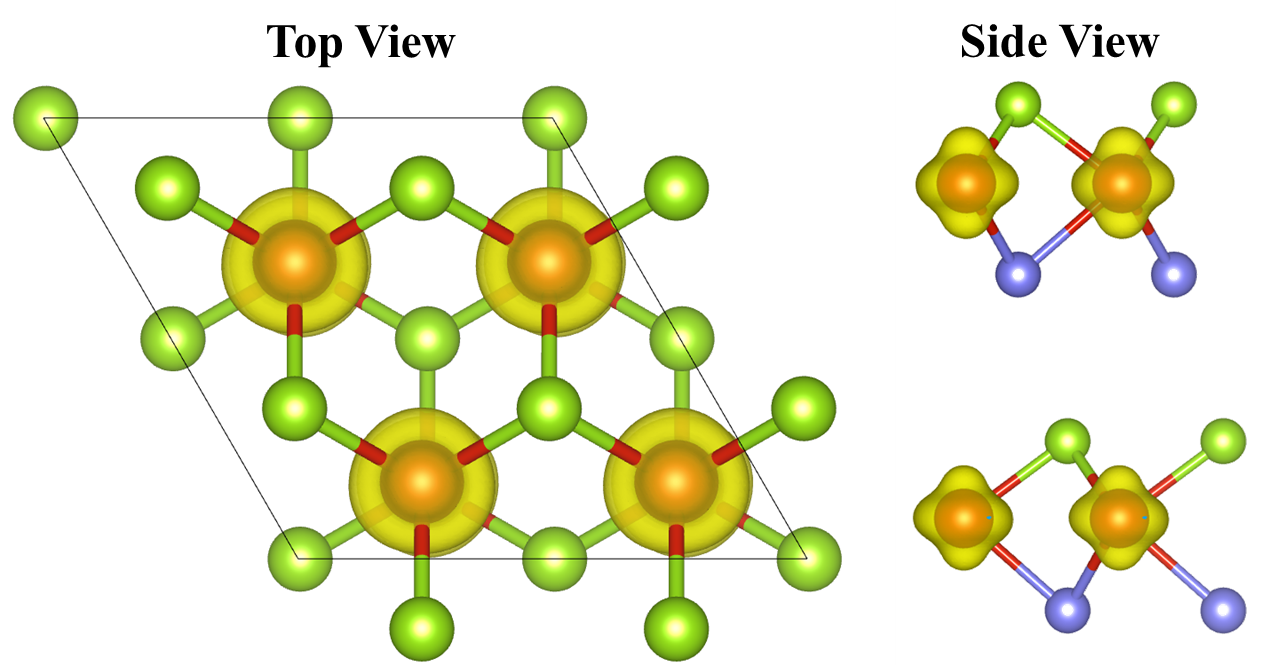} 
\par\end{centering}
\caption{The calculated spin density difference for the VSSe bilayer under
consideration, plotted at an isosurface value of 0.0186 e/{\AA$^{3}$}.
The yellow color represents the spin density corresponding to majority
spin carriers.}
\label{fig:spin-density} 
\end{figure}

The magnetic moments associated with each atomic species are computed
using the GGA+\textit{U} method, and their calculated values (see
Table \ref{tab:magnetic-moment}) for the two V atoms, V1 and V2 in
the bilayer are 1.18 $\mu_{B}$ and 1.19 $\mu_{B}$, respectively,
which are almost identical to 1.18 $\mu_{B}$ computed for the monolayer.
As anticipated, the primary contribution to magnetism comes from the
transition metal atoms (V) of both the layers due to the unpaired
electrons in \textit{d} orbitals. Further, we computed the spin density
difference defined as $M(\overrightarrow{r})=\rho_{\uparrow}(\overrightarrow{r})-\rho_{\downarrow}(\overrightarrow{r})$,
where $\rho_{\uparrow}(\overrightarrow{r})$ and $\rho_{\downarrow}(\overrightarrow{r})$
signify the majority and minority spin densities, respectively, to
understand the distribution of the magnetic moments on different atoms.
From the $M(\vec{r})$ plotted in Fig. \ref{fig:spin-density} for
our considered FM VSSe bilayer, it is clear that the spin density,
which is contributed by the majority spin carriers, primarily resides
on the vanadium atoms, consistent with their significant contribution
to the overall magnetic moment. The ferromagnetism observed within
the intra- and interlayer domains of the VSSe bilayer can be attributed
to the indirect exchange coupling between the magnetic moments localized
on vanadium, termed super-exchange (SE) \cite{Dey2020}. Intralayer
SE arises from V-S-V and V-Se-V coupling, while the interlayer SE
stems from V-Se-S-V couplings. Further, the bond angles in the top
and bottom layers of the VSSe bilayer are close to 90$^{\circ}$,
confirming the FM superexchange in the bilayer according to the Goodenough-Kanamoris
rule \cite{goodenough2008}. Similar superexchange-based couplings
have already been reported for the case of the monolayer \cite{Dey2020}.
Additionally, magnetic anisotropy calculations are performed to ascertain
the easy-plane/axis of magnetization in the considered FM Janus VSSe
bilayer, because the magnetic anisotropy significantly influences
the stability of 2D ferromagnets. For example, Dey \textit{et al.}
\cite{Dey2020} reported the easy-plane of magnetization to be the
plane of the VSSe monolayer, with the spins oriented along it. We
computed the magnetic anisotropy energy (MAE) as MAE $=E_{[100]}-E_{[001]}$,
where $E_{[100]}$ and $E_{[001]}$ denote the energies corresponding
to the in-plane and out-of-plane magnetization orientations, respectively.
For the purpose, we employed the GGA+\textit{U} approach, including
the spin-orbit coupling to account for the relativistic effects, and
obtained MAE to be -0.66 meV indicating that the FM Janus VSSe bilayer
also manifests an easy plane of magnetization, which is the plane
of the bilayer. This suggests that the intrinsic magnetic anisotropy
of the VSSe monolayer persists even after bilayer formation, due to
the weak nature of the vdW interaction holding the two layers together.
Moreover, our MAE value for the VSSe bilayer is quite close to the
corresponding value for the monolayer \cite{Dey2020}.

\begin{table}[H]
\caption{Localized magnetic moments (in the units of Bohr magneton, $\mu_{B}$)
on each of the constituent elements. }
\label{tab:magnetic-moment} \centering{}%
\begin{tabular}{ccccccc}
\toprule 
\multirow{1}{*}{Method} & \multicolumn{3}{c}{Layer 1} & \multicolumn{3}{c}{Layer 2}\tabularnewline
 & V1  & S1  & Se1  & V2  & S2  & Se2 \tabularnewline
\midrule 
GGA+\textit{U}  & 1.18  & -0.08  & -0.13  & 1.19  & -0.08  & -0.14\tabularnewline
\bottomrule
\end{tabular}
\end{table}

\subsection{Effect of external electric field on the electronic properties}

\begin{figure}[H]
\begin{centering}
\includegraphics[scale=0.3]{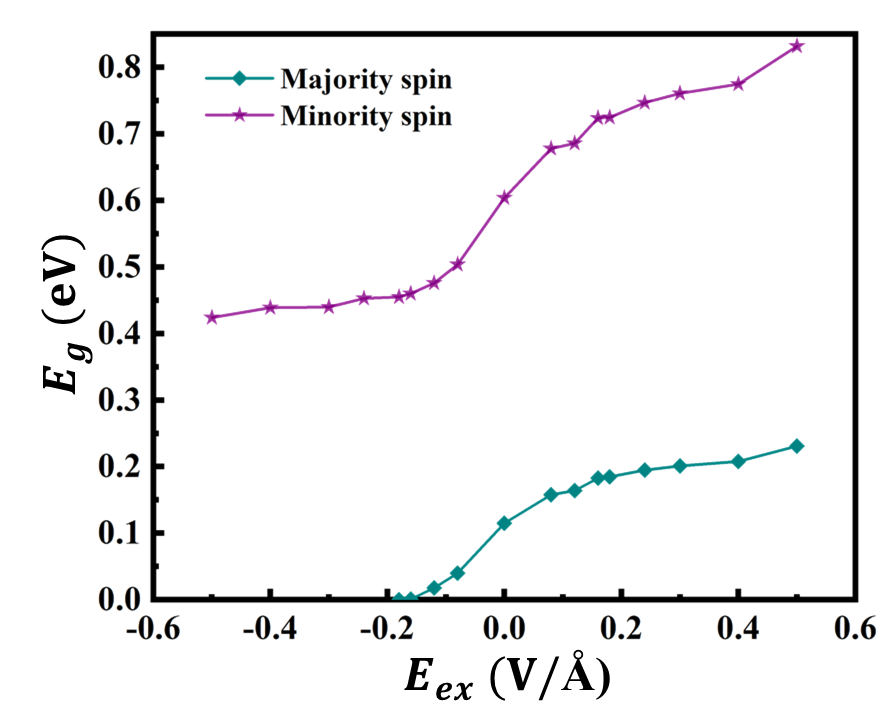} 
\par\end{centering}
\caption{Variation of $E_{g}$ with the strengths of applied $\mathbf{E_{ex}}$.
The positive (negative) values of $\mathbf{E_{ex}}$ imply it is in
the upward (downward) direction, normal to the plane of considered
VSSe bilayer. }
\label{fig:gap-polariz} 
\end{figure}

This section elucidates the effect of external electric field on the
electronic properties of the FM Janus VSSe bilayer. As discussed in
Sec. \ref{subsec:Electronic-and-magnetic}, the energy states close
to $E_{F}$ in both the valence band and conduction band regions correspond
to the majority spin channel suggesting the possibility of gap closing
under an external perturbation. This closing of one spin channel can
lead to half-metallic behavior which can in turn be useful for applications
in spintronics. Here we use an external electric field ($\mathbf{E_{ex}}$)
as a perturbation, to see the effect on the electronic structure and
band gap of the VSSe bilayer. $\mathbf{E_{ex}}$ is applied normal
to the considered VSSe bilayer with its magnitude ranging from 0.08
V/$\text{\AA}$ to 0.50 V/$\text{\AA}$, in both the upward and downward
directions, denoted by $\mathbf{E_{ex}^{\uparrow}}$ and $\mathbf{E_{ex}^{\downarrow}}$,
respectively. We observe that for both spin channels, $E_{g}$ increases
with increasing strengths of $\mathbf{E_{ex}^{\uparrow}}$, while,
decreasing behavior in $E_{g}$ is observed with increasing $\mathbf{E_{ex}^{\downarrow}}$.
This observed change in $E_{g}$ with direction and intensities of
$\mathbf{E_{ex}}$ is depicted in Fig. \ref{fig:gap-polariz}. The
VSSe bilayer with the electric field applied upward can be useful
in applications that demand narrow band gap semiconductors, such as
adaptive radiation sensors and tunable IR photodetectors \cite{Zha2022}.
To investigate the effect of applied electric field on charge distribution,
we plotted charge density difference for with and without $\mathbf{E_{ex}}$
case, i.e., $\rho(\boldsymbol{E_{ex}})-\rho(0)$. The difference is
plotted for an electric field of strength 0.18 V/{\AA} applied
in both the upward and downward directions, as shown in Fig. S7 of
the SM \cite{supporting-data}. It is observed that the charges accumulate
at the top and bottom layers of the bilayer on applying $\mathbf{E_{ex}}$
in the upward (Fig. S7(a)) and downward (see Fig. S7(b)) directions,
respectively.

\begin{figure}[H]
\begin{centering}
\includegraphics[scale=0.4]{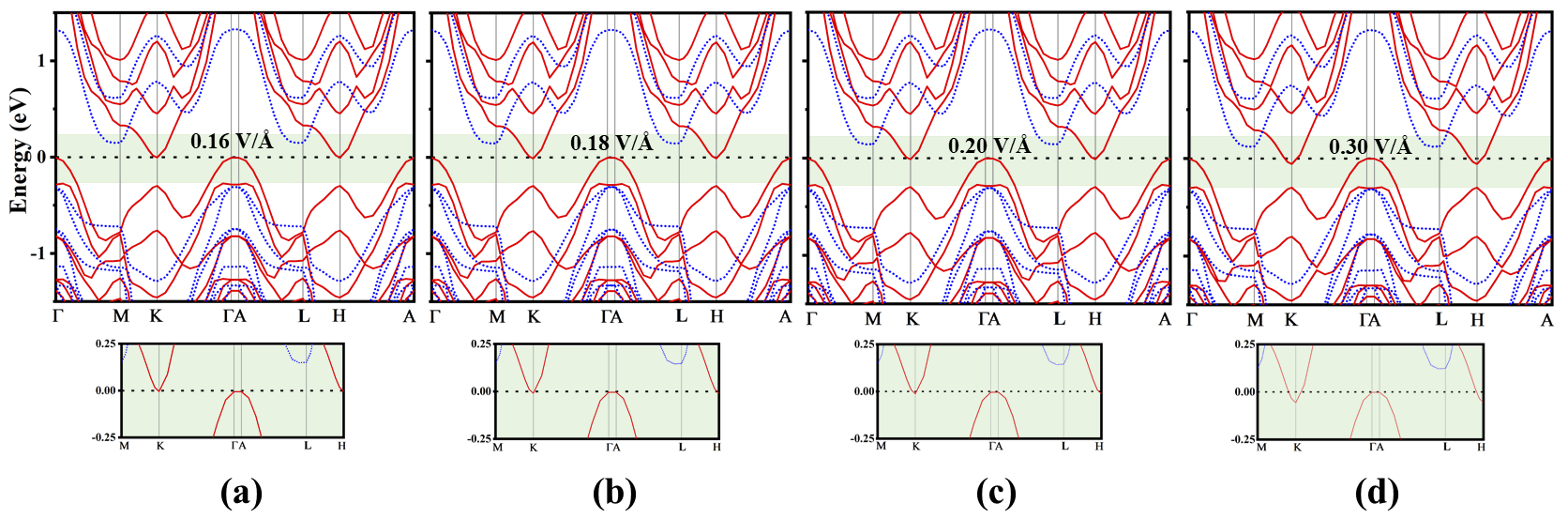} 
\par\end{centering}
\caption{Spin-polarized band structures of the VSSe bilayer in the presence
of four different $\boldsymbol{\mathbf{E_{ex}^{\downarrow}}}$ strengths.
The dashed line at zero represents $E_{F}$. }
\label{fig:EF-ng-bandstrc} 
\end{figure}

We further explore the effect of $\mathbf{E_{ex}^{\downarrow}}$,
as it leads to the lowering of the gap. The band structures for four
varying strengths of $\mathbf{E_{ex}^{\downarrow}}$ are presented
in Fig. \ref{fig:EF-ng-bandstrc}, from where it is obvious that the
application of $\mathbf{E_{ex}^{\downarrow}}$ results in gradual
downward and upward shifts of the CBM and VBM, respectively, for both
spin channels. However, the \textbf{k}-point corresponding to the
VBM and CBM for both spins remains unchanged upon applying $\mathbf{E_{ex}^{\downarrow}}$.
The fundamental gap closes ($E_{g}\rightarrow0$) for the majority
spin carriers at $\mathbf{E_{ex}^{\downarrow}}$ = 0.16 V/{\AA}
(Fig. \ref{fig:EF-ng-bandstrc}(a)), with the CBM and VBM of the majority
spin channel just touching each other, while a significant gap still
exists for the minority spin carriers. Therefore, our calculations
predict the considered FM Janus VSSe bilayer to be a spin gapless
semiconductor (SGS) at that field strength ($\mathbf{E_{ex}^{\downarrow}}$
= 0.16 V/{\AA}). SGS is a subclass of conventional half-metallic
ferromagnets, providing a new route to spintronics \cite{Wang2008,Yue2020}.
When $\mathbf{E_{ex}^{\downarrow}}$ is increased to 0.18 V/{\AA}
(Fig. \ref{fig:EF-ng-bandstrc}(b)), the CBM and VBM of majority spin
crosses each other at $E_{F}$ and a small overlap of about 9 meV
between them (CBM and VBM) is observed. The overlap between the CBM
and VBM corresponding to the majority spin channel increases further
with increasing $\mathbf{E_{ex}^{\downarrow}}$ ($\geq$ 0.2 V/$\text{\AA}$),
while a significant $E_{g}$ still exists for the minority spin channel
(see Fig. \ref{fig:EF-ng-bandstrc}(c) and (d)). Therefore, a crossover
from semiconducting to half-metallic behavior of the considered VSSe
bilayer is achieved for $\mathbf{E_{ex}^{\downarrow}}$ $\geq$ 0.18
V/{\AA}. This value of the required electric field is lower than
the values reported for other systems to undergo semiconductor-to-half-metal
transition. For instance, the half-metallic behavior reported for
CrI$_{3}$-CrGeTe$_{3}$ heterobilayer is at an external field of
0.6 V/{\AA} \cite{Tang2020}, while in the CrSBr monolayer, it
has been reported at a field strength of 0.3 -- 0.4 V/{\AA} \cite{Guo2023}.
Notably, the predicted electric field strength for the VSSe bilayer
($\approx$ 0.18 V/{\AA}) is both smaller and practically feasible,
as a study by Benjamin \textit{et al.} \cite{Weintrub2022} demonstrated
the realization of an intense electric field larger than 0.4 V/{\AA}
using dual ionic grating.

The resulting spin gapless ($\mathbf{E_{ex}^{\downarrow}}=$ 0.16
V/{\AA}) and half-metallic ($\mathbf{E_{ex}^{\downarrow}}$ $\geq$
0.18 V/{\AA}) behaviors can also be verified from the corresponding
value of TDOS for the majority spin carriers at $E_{F}$. A small
value of TDOS (0.33 eV$^{-1}$) at $E_{F}$ favors the spin gapless
nature for the case of $\mathbf{E_{ex}^{\downarrow}}$ equaling 0.16
V/{\AA}. With the increasing $\mathbf{E_{ex}^{\downarrow}}$, a
rise in the TDOS at $E_{F}$ is observed. For the applied fields of
0.18 V/{\AA}, 0.20 V/{\AA}, and 0.30 V/{\AA}, enhanced TDOS
of magnitudes 0.40, 0.44, and 0.64 eV$^{-1}$ at $E_{F}$ are obtained,
respectively (See Fig. S8 of the SM for TDOS plots). For the resulting
half-metallic VSSe bilayer (case of $\mathbf{E_{ex}^{\downarrow}}=$
0.18 V/{\AA}), the intrinsic spin polarization ($P_{s}$) is calculated
using the following formula 
\begin{equation}
P_{s}=\frac{n_{\uparrow}(E_{F})-n_{\downarrow}(E_{F})}{n_{\uparrow}(E_{F})+n_{\downarrow}(E_{F})}\times100
\end{equation}

where $n_{\uparrow}(E_{F})$ and $n_{\downarrow}(E_{F})$ represent
the density of states at $E_{F}$ for the majority and minority spin
channels, respectively. Due to zero $n_{\downarrow}(E_{F})$, 100
\% intrinsic spin polarization is obtained which suggests that the
VSSe bilayer can be explored for spintronics. It is worth mentioning
that we also investigated the effect of external electric field on
the band gap of VSSe monolayer, but no such change in the band gap
is observed, clearly implying that the spin-gapless as well as half-metallic
behaviors discussed above are unique properties of the bilayer.

\begin{figure}[H]
\begin{centering}
\includegraphics[scale=0.45]{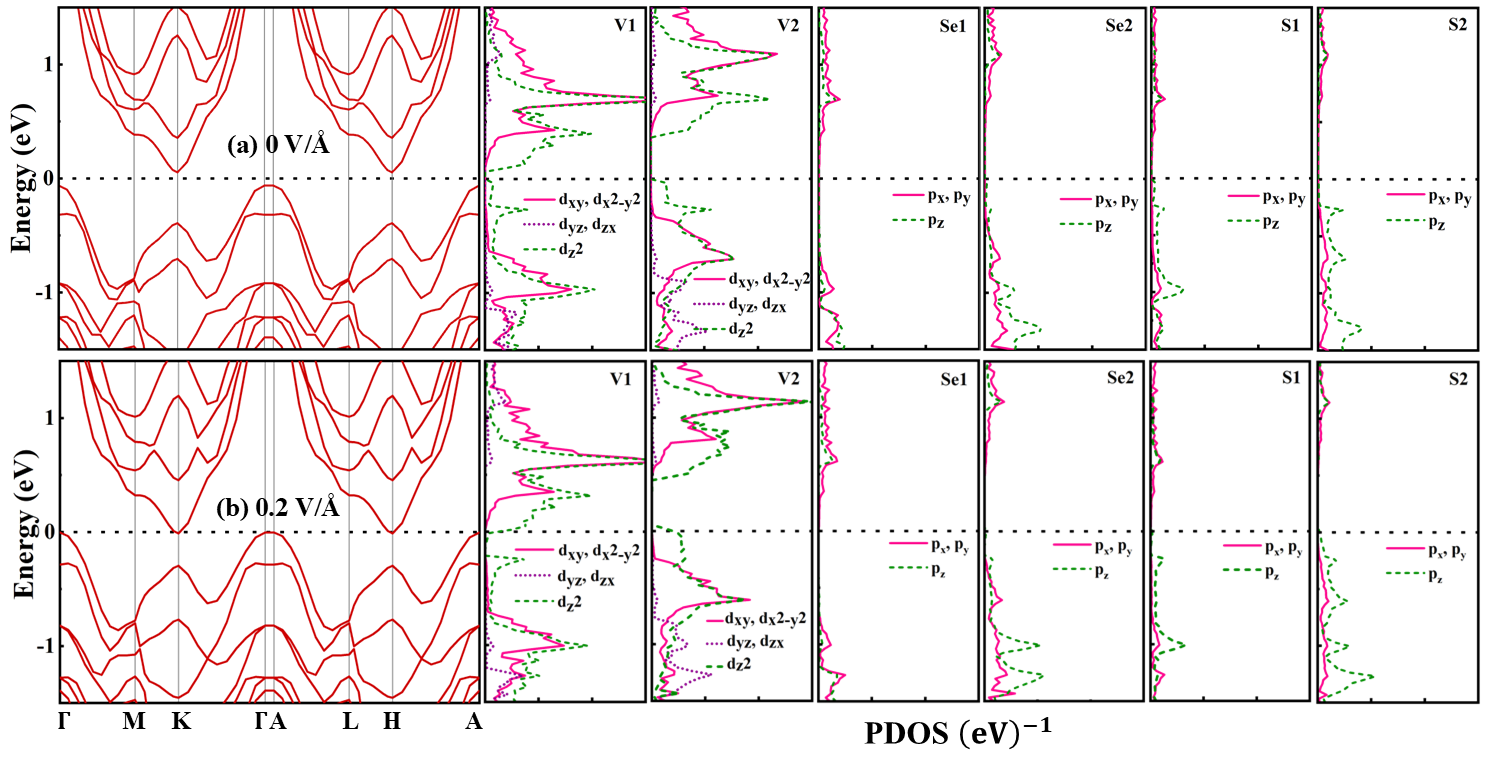} 
\par\end{centering}
\caption{Band structure and \textit{lm}-decomposed atom projected PDOS corresponding
to the majority spin carriers of the VSSe bilayer when (a) $\mathbf{E_{ex}}$
is absent and (b) $\mathbf{E_{ex}^{\downarrow}}$ = 0.2 V/{\AA}.
The dashed line at zero represents $E_{F}$. }
\label{fig:lmpdos} 
\end{figure}

\begin{figure}[H]
\begin{centering}
\includegraphics[scale=0.5]{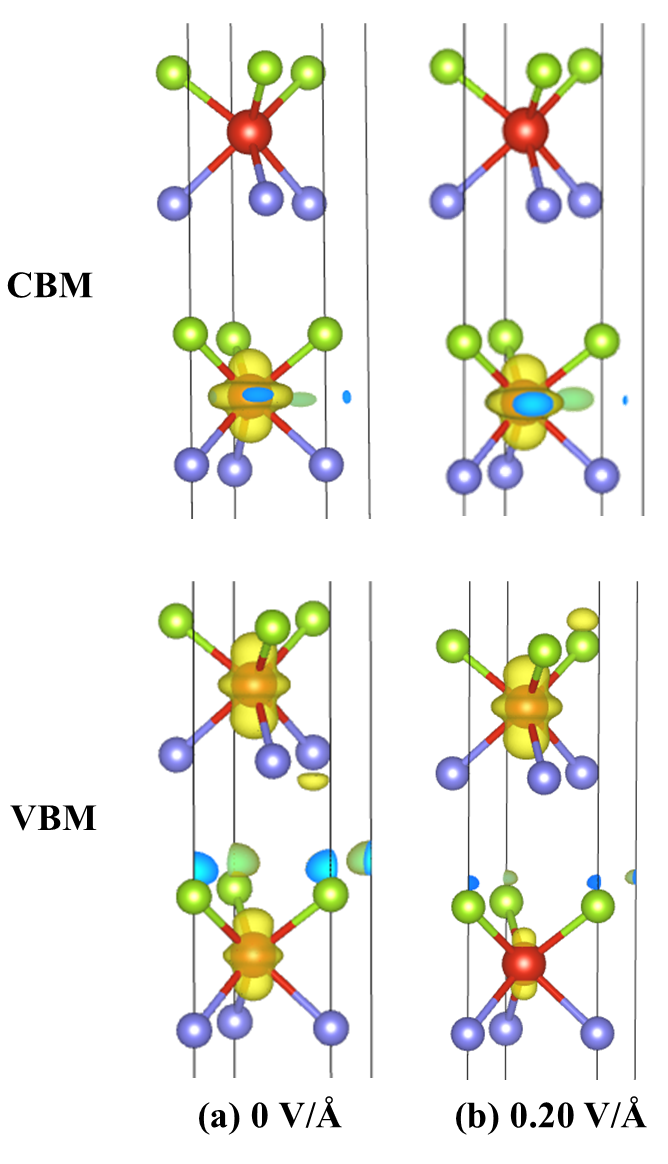} 
\par\end{centering}
\caption{Charge density composition of the VBM and CBM corresponding to the
majority spin carriers of the VSSe bilayer for cases: (a) $\mathbf{E_{ex}}$
is absent and (b) $\mathbf{E_{ex}^{\downarrow}}$ = 0.2 V/{\AA}.
All the plots are at an isosurface value of 0.016 e/{\AA$^{3}$}.\textcolor{red}{{}
}The yellow and blue colors depict the charge accumulation and depletion
regions, respectively.}
\label{fig:vbm-cbm-chg} 
\end{figure}

To get a picture of the role of atomic orbitals in gap closing, \textit{l}
and \textit{m} (azimuthal and magnetic quantum numbers) decomposed
atom-projected PDOS are plotted corresponding to the majority spin
carriers. The following two different cases corresponding to $\mathbf{E_{ex}^{\downarrow}}$:
(a) no external electric field and (b) at $\mathbf{E_{ex}^{\downarrow}}$
= 0.20 V/{\AA} are considered, as depicted in Fig. \ref{fig:lmpdos}
along with the respective band structures. The non-identical PDOS
signatures of the two V atoms (V1 and V2), two S atoms (S1 and S2),
and two Se atoms (Se1 and Se2) are due to broken horizontal mirror
symmetry. For the pristine VSSe bilayer (Fig. \ref{fig:lmpdos}(a)),
VBM is composed primarily of the $d_{z^{2}}$ orbital of V2 atom of
top layer with a small contribution of the same orbital of V1 of bottom
layer. Also, in the CBM, there is a small mixing of V1 $d_{xy},d_{x^{2}-y^{2}}$
orbitals with the $d_{z^{2}}$ orbital. Additionally, $p_{z}$ orbitals
of S1 and Se2 atoms make small contributions to VBM, whereas, a negligible
contribution comes from the Se1 and S2 atoms as they do not interact
with the $p_{z}$ orbitals of the neighboring layers due to the presence
of vacuum. With the increasing magnitude of $\mathbf{E_{ex}^{\downarrow}},$
the valence (conduction) band states exhibit upward (downward) shifts
leading to the reduction of the band gap, and consequently, half metallicity
as discussed earlier. The changes in the contribution from atomic
orbitals for $\mathbf{E_{ex}^{\downarrow}}$ = 0.20 V/{\AA}, compared
to the $\mathbf{E_{ex}^{\downarrow}}$= 0 case, can be observed in
the PDOS plot presented in Fig. \ref{fig:lmpdos}(b). The redistribution
of states of $d_{z^{2}}$ orbitals around $E_{F}$ indicates that
the $d_{z^{2}}$ orbitals of both the V atoms (V1 from CBM side and
V2 from VBM side) primarily contribute to the closing of the gap.
These results of the \textit{lm}-decomposed PDOS are also reflected
in the band decomposed charge density distribution plotted for the
CBM and VBM corresponding to the majority spin carriers in Fig. \ref{fig:vbm-cbm-chg}.
Fig. \ref{fig:vbm-cbm-chg} suggests: (a) in the absence of the external
field, the contribution to CBM comes completely from the bottom layer,
whereas, the top layer mainly contributes to the VBM with small participation
from the bottom layer, and (b) at $\mathbf{E_{ex}}$ = 0.2 V/{\AA},
the VBM and CBM are mainly distributed over the top and bottom layers,
respectively.

\begin{figure}
\begin{centering}
\includegraphics[scale=0.4]{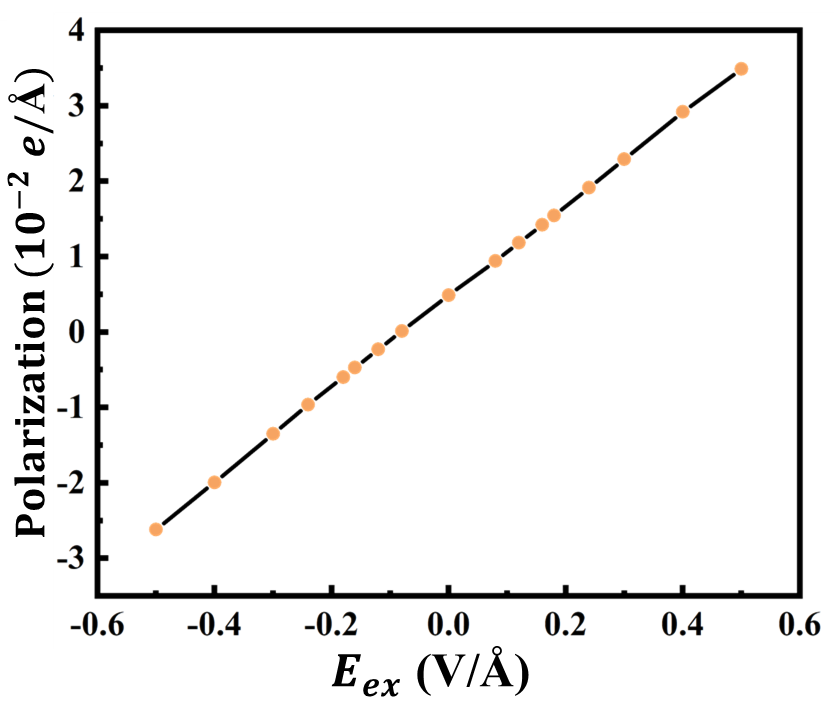} 
\par\end{centering}
\caption{Electric polarization ($\mathbf{P_{ex}}$) plotted as a function of
$\mathbf{E_{ex}}$. The positive (negative) values of $\mathbf{E_{ex}}$
imply it is in the upward (downward) direction, normal to the plane
of considered VSSe bilayer.}
\label{fig:EF_polarization} 
\end{figure}

In order to understand the static dielectric response of the VSSe
bilayer, we calculated its electric polarization, $\mathbf{P_{ex}}$,
for different values of the applied electric field, and the results
are plotted in Fig. \ref{fig:EF_polarization}. A significant non-zero
value of $\mathbf{P_{ex}}=$ 0.49$\times10^{-2}e/\text{\AA}$ at $\mathbf{E_{ex}}$
= 0, shows the spontaneous polarization which arises due to $\mathbf{E_{in}}$
caused by the lack of mirror symmetry, as discussed earlier. Upon
application of $\mathbf{E_{ex}}$, polarization increases monotonically
with the strength of $\mathbf{E_{ex}^{\uparrow}}$, whereas, it decreases
with the increase in strength of $\mathbf{E_{ex}^{\downarrow}}$.
The reason for former is that $\mathbf{E_{ex}^{\uparrow}}$ is applied
along the direction of $\mathbf{E_{in}}$, while in the latter case
the direction of $\mathbf{E_{ex}^{\downarrow}}$ is opposite to that
of $\mathbf{E_{in}}$. From these observations we infer that $\mathbf{E_{ex}^{\uparrow}}$
strengthens the intrinsic dipole moment present in the bilayer structure,
thereby increasing the corresponding polarization. However, $\mathbf{E_{ex}^{\downarrow}}$
attempts to reorient the intrinsic dipole in the opposite direction,
i.e., in the direction of $\mathbf{E_{ex}^{\downarrow}}$. This initially
decreases the magnitude of the dipole moment and polarization from
the intrinsic one, and then it increases in the opposite direction
once the dipole aligns in this direction. Furthermore, the $\mathbf{P_{ex}-\mathbf{E_{ex}}}$
plot of Fig. \ref{fig:EF_polarization} is strictly linear, implying
that the VSSe bilayer considered in this work behaves like a linear
dielectric.

\section{conclusion}

In this work we explored the effects of external electric field on
the electronic band structure of vanadium-based 2D Janus VSSe bilayer
and its potential application in spintronics. The structural, mechanical,
electronic and magnetic properties of the bilayer have been investigated.
The study shows that the VSSe bilayer is a ferromagnetic semiconductor
possessing an easy-plane of magnetization when stacked in AB configuration
with indirect band gaps of 0.11 eV and 0.60 eV for majority and minority
spin carriers, respectively. When subjected to an external electric
field perpendicular to the plane, applied in both upward and downward
directions, the bilayer exhibits tunable electronic properties. Our
findings reveal that the band gap decreases (increases) with the increasing
external field downward (upward). Notably, a semiconductor to half-metal
transition occurs at a field strength of $\approx$ 0.18 V/{\AA},
applied downward, while a spin gapless semiconducting behavior is
observed at 0.16 V/{\AA}. These transitions to half-metallic and
spin gapless states are driven by the closing of the band gap corresponding
to the majority spin, primarily due to the interaction of the vanadium
atom's out-of-plane $d_{z^{2}}$ orbitals across both layers. Furthermore,
VSSe bilayer exposed to an upward electric field can be used in radiation
sensors and IR photodetectors for which narrow band gap semiconductors
are well suited. The ability to tune the band gap and obtain semiconductor-to-half-metal
transition or vice-versa by reversing the electric field direction
demonstrates the versatility of the VSSe bilayer. Our study predicts
Janus VSSe bilayer as a promising candidate for future spintronic
devices.

\section*{acknowledgments}

One of the authors, K.D. acknowledges financial assistance from the
Prime Minister Research Fellowship (PMRF ID-1302054), MHRD, India,
and Space Time computational facility of Indian Institute of Technology
Bombay. S.S.S. acknowledges the support through the Institute Post-Doctoral
Fellowship (IPDF) of Indian Institute of Technology Bombay.

\section*{data availability statement}

Majority of the data is given in the main manuscript and Supplemental
Material and the rest of the data will be available from the authors
upon reasonable request.

 \bibliographystyle{unsrt}
\bibliography{Manuscript}

\begin{thebibliography}{10}

\bibitem{Ahn2020}
Ethan~C. Ahn.
\newblock {2D materials for spintronic devices}.
\newblock {\em npj 2D Materials and Applications}, 4(1):17, Jun 2020.

\bibitem{cortie2020}
David~L. Cortie, Grace~L. Causer, Kirrily~C. Rule, Helmut Fritzsche, Wolfgang
  Kreuzpaintner, and Frank Klose.
\newblock {Two-Dimensional Magnets: Forgotten History and Recent Progress
  towards Spintronic Applications}.
\newblock {\em Advanced Functional Materials}, 30(18):1901414, 2020.

\bibitem{Liu2019}
Xiaolong Liu and Mark~C. Hersam.
\newblock {2D materials for quantum information science}.
\newblock {\em Nature Reviews Materials}, 4(10):669--684, Oct 2019.

\bibitem{han2023}
Fanjunjie Han, Xu~Yan, Fei Li, Hong Yu, Wenjing Li, Xin Zhong, Aitor Bergara,
  and Guochun Yang.
\newblock {Prediction of monolayer $FeP_{4}$ with intrinsic half-metal
  ferrimagnetism above room temperature}.
\newblock {\em Phys. Rev. B}, 107:024414, Jan 2023.

\bibitem{zhang2020}
Dechen Zhang, Azizur Rahman, Wei Qin, Xingxing Li, Ping Cui, Zengming Zhang,
  and Zhenyu Zhang.
\newblock {Prediction of ${\mathrm{MnSiTe}}_{3}$ as an intrinsic layered
  half-metal}.
\newblock {\em Phys. Rev. B}, 101:205119, May 2020.

\bibitem{Yao2021}
Qiushi Yao, Jiayu Li, and Qihang Liu.
\newblock {Fragile symmetry-protected half metallicity in two-dimensional van
  der Waals magnets: A case study of monolayer $\mathrm{Fe}{\mathrm{Cl}}_{2}$}.
\newblock {\em Phys. Rev. B}, 104:035108, Jul 2021.

\bibitem{Wu2018}
Qisheng Wu, Yehui Zhang, Qionghua Zhou, Jinlan Wang, and Xiao~Cheng Zeng.
\newblock {Transition-Metal Dihydride Monolayers: A New Family of
  Two-Dimensional Ferromagnetic Materials with Intrinsic Room-Temperature
  Half-Metallicity}.
\newblock {\em The Journal of Physical Chemistry Letters}, 9(15):4260--4266,
  Aug 2018.

\bibitem{Ashton2017}
Michael Ashton, Dorde Gluhovic, Susan~B. Sinnott, Jing Guo, Derek~A. Stewart,
  and Richard~G. Hennig.
\newblock {Two-Dimensional Intrinsic Half-Metals With Large Spin Gaps}.
\newblock {\em Nano Letters}, 17(9):5251--5257, Sep 2017.

\bibitem{Wang2019}
Bing Wang, Yehui Zhang, Liang Ma, Qisheng Wu, Yilv Guo, Xiwen Zhang, and Jinlan
  Wang.
\newblock {MnX (X = P{,} As) monolayers: a new type of two-dimensional
  intrinsic room temperature ferromagnetic half-metallic material with large
  magnetic anisotropy}.
\newblock {\em Nanoscale}, 11:4204--4209, 2019.

\bibitem{Pei2018}
Qi~Pei, Xiaocha Wang, Jijun Zou, and Wenbo Mi.
\newblock {Half-metallicity and spin-valley coupling in 5d transition metal
  substituted monolayer $MnPSe_{3}$}.
\newblock {\em J. Mater. Chem. C}, 6:8092--8098, 2018.

\bibitem{Scalise2012}
Emilio Scalise, Michel Houssa, Geoffrey Pourtois, Valery Afanas'ev, and
  Andr{\'e} Stesmans.
\newblock {Strain-induced semiconductor to metal transition in the
  two-dimensional honeycomb structure of $MoS_{2}$}.
\newblock {\em Nano Research}, 5(1):43--48, Jan 2012.

\bibitem{Rama2011}
Ashwin Ramasubramaniam, Doron Naveh, and Elias Towe.
\newblock Tunable band gaps in bilayer transition-metal dichalcogenides.
\newblock {\em Phys. Rev. B}, 84:205325, Nov 2011.

\bibitem{Yang_2020}
Qiang Yang, Liangzhi Kou, Xiaohui Hu, Yifeng Wang, Chunhua Lu, Arkady~V
  Krasheninnikov, and Litao Sun.
\newblock {Strain robust spin gapless semiconductors/half-metals in transition
  metal embedded $MoSe_{2}$ monolayer}.
\newblock {\em Journal of Physics: Condensed Matter}, 32(36):365305, jun 2020.

\bibitem{Yang2010}
Zailin Yang and Jun Ni.
\newblock {Modulation of electronic properties of hexagonal boron nitride
  bilayers by an electric field: A first principles study}.
\newblock {\em Journal of Applied Physics}, 107(10):104301, 05 2010.

\bibitem{imt-zno-2018}
Le~Zhang, Shanshan Chen, Xiangyang Chen, Zhizhen Ye, and Liping Zhu.
\newblock {Electric-field driven insulator-metal transition and tunable
  magnetoresistance in ZnO thin film}.
\newblock {\em Applied Physics Letters}, 112(15):153505, 04 2018.

\bibitem{San-Dong1}
Yichen Liu, San-Dong Guo, Yongpan Li, and Cheng-Cheng Liu.
\newblock {Two-Dimensional Fully Compensated Ferrimagnetism}.
\newblock {\em Phys. Rev. Lett.}, 134:116703, Mar 2025.

\bibitem{San-Dong2}
San-Dong Guo.
\newblock Valley polarization in two-dimensional zero-net-magnetization
  magnets.
\newblock {\em Applied Physics Letters}, 126(8):080502, 02 2025.

\bibitem{Guo2023}
Hao-Tian Guo, San-Dong Guo, and Yee~Sin Ang.
\newblock {Electric-field induced half-metallic properties in an experimentally
  synthesized CrSBr monolayer}.
\newblock {\em Phys. Chem. Chem. Phys.}, 25:30269--30275, 2023.

\bibitem{Tang2020}
Cheng Tang, Lei Zhang, Stefano Sanvito, and Aijun Du.
\newblock Electric-controlled half-metallicity in magnetic van der waals
  heterobilayer.
\newblock {\em J. Mater. Chem. C}, 8:7034--7040, 2020.

\bibitem{Duan2015}
Xidong Duan, Chen Wang, Anlian Pan, Ruqin Yu, and Xiangfeng Duan.
\newblock Two-dimensional transition metal dichalcogenides as atomically thin
  semiconductors: opportunities and challenges.
\newblock {\em Chem. Soc. Rev.}, 44:8859--8876, 2015.

\bibitem{zhuang2016}
Houlong~L. Zhuang and Richard~G. Hennig.
\newblock {Stability and magnetism of strongly correlated single-layer
  ${\mathrm{VS}}_{2}$}.
\newblock {\em Phys. Rev. B}, 93:054429, Feb 2016.

\bibitem{Ma2012}
Yandong Ma, Ying Dai, Meng Guo, Chengwang Niu, Yingtao Zhu, and Baibiao Huang.
\newblock {Evidence of the Existence of Magnetism in Pristine $VX_{2}$
  Monolayers (X = S, Se) and Their Strain-Induced Tunable Magnetic Properties}.
\newblock {\em ACS Nano}, 6(2):1695--1701, Feb 2012.

\bibitem{chen2020}
Wei Chen, Jian-min Zhang, Yao-zhuang Nie, Qing-lin Xia, and Guang-hua Guo.
\newblock {Electronic structure and magnetism of $MTe_{2}$ (M= Ti, V, Cr, Mn,
  Fe, Co and Ni) monolayers}.
\newblock {\em Journal of Magnetism and Magnetic Materials}, 508:166878, 2020.

\bibitem{mermin1996}
N.~D. Mermin and H.~Wagner.
\newblock {Absence of Ferromagnetism or Antiferromagnetism in One- or
  Two-Dimensional Isotropic Heisenberg Models}.
\newblock {\em Phys. Rev. Lett.}, 17:1133--1136, Nov 1966.

\bibitem{Huang2017}
Bevin Huang, Genevieve Clark, Efr{\'e}n Navarro-Moratalla, Dahlia~R. Klein, Ran
  Cheng, Kyle~L. Seyler, Ding Zhong, Emma Schmidgall, Michael~A. McGuire,
  David~H. Cobden, Wang Yao, Di~Xiao, Pablo Jarillo-Herrero, and Xiaodong Xu.
\newblock Layer-dependent ferromagnetism in a van der waals crystal down to the
  monolayer limit.
\newblock {\em Nature}, 546(7657):270--273, Jun 2017.

\bibitem{Gong2017}
Cheng Gong, Lin Li, Zhenglu Li, Huiwen Ji, Alex Stern, Yang Xia, Ting Cao, Wei
  Bao, Chenzhe Wang, Yuan Wang, Z.~Q. Qiu, R.~J. Cava, Steven~G. Louie, Jing
  Xia, and Xiang Zhang.
\newblock Discovery of intrinsic ferromagnetism in two-dimensional van der
  waals crystals.
\newblock {\em Nature}, 546(7657):265--269, Jun 2017.

\bibitem{OHara2018}
Dante~J. O'Hara, Tiancong Zhu, Amanda~H. Trout, Adam~S. Ahmed, Yunqiu~Kelly
  Luo, Choong~Hee Lee, Mark~R. Brenner, Siddharth Rajan, Jay~A. Gupta, David~W.
  McComb, and Roland~K. Kawakami.
\newblock {Room Temperature Intrinsic Ferromagnetism in Epitaxial Manganese
  Selenide Films in the Monolayer Limit}.
\newblock {\em Nano Letters}, 18(5):3125--3131, May 2018.

\bibitem{Bonilla2018}
Manuel Bonilla, Sadhu Kolekar, Yujing Ma, Horacio~Coy Diaz, Vijaysankar
  Kalappattil, Raja Das, Tatiana Eggers, Humberto~R. Gutierrez, Manh-Huong
  Phan, and Matthias Batzill.
\newblock {Strong room-temperature ferromagnetism in $VSe_{2}$ monolayers on
  van der Waals substrates}.
\newblock {\em Nature Nanotechnology}, 13(4):289--293, Apr 2018.

\bibitem{Hu2018}
Tao Hu, Fanhao Jia, Guodong Zhao, Jiongyao Wu, Alessandro Stroppa, and Wei Ren.
\newblock {Intrinsic and anisotropic Rashba spin splitting in Janus
  transition-metal dichalcogenide monolayers}.
\newblock {\em Phys. Rev. B}, 97:235404, Jun 2018.

\bibitem{Riis-Jensen2019}
Anders~C. Riis-Jensen, Thorsten Deilmann, Thomas Olsen, and Kristian~S.
  Thygesen.
\newblock {Classifying the Electronic and Optical Properties of Janus
  Monolayers}.
\newblock {\em ACS Nano}, 13(11):13354--13364, Nov 2019.

\bibitem{Zhang2019}
Chunmei Zhang, Yihan Nie, Stefano Sanvito, and Aijun Du.
\newblock {First-Principles Prediction of a Room-Temperature Ferromagnetic
  Janus VSSe Monolayer with Piezoelectricity, Ferroelasticity, and Large Valley
  Polarization}.
\newblock {\em Nano Letters}, 19(2):1366--1370, Feb 2019.

\bibitem{Zhang2017}
Jing Zhang, Shuai Jia, Iskandar Kholmanov, Liang Dong, Dequan Er, Weibing Chen,
  Hua Guo, Zehua Jin, Vivek~B. Shenoy, Li~Shi, and Jun Lou.
\newblock {Janus Monolayer Transition-Metal Dichalcogenides}.
\newblock {\em ACS Nano}, 11(8):8192--8198, Aug 2017.

\bibitem{junjie2018}
Junjie He and Shuo Li.
\newblock {Two-dimensional Janus transition-metal dichalcogenides with
  intrinsic ferromagnetism and half-metallicity}.
\newblock {\em Computational Materials Science}, 152:151--157, 2018.

\bibitem{Dey2020}
Dibyendu Dey and Antia~S. Botana.
\newblock {Structural, electronic, and magnetic properties of vanadium-based
  Janus dichalcogenide monolayers: A first-principles study}.
\newblock {\em Phys. Rev. Mater.}, 4:074002, Jul 2020.

\bibitem{Luo2020}
Chaobo Luo, Xiangyang Peng, Jinfeng Qu, and Jianxin Zhong.
\newblock {Valley degree of freedom in ferromagnetic Janus monolayer H-VSSe and
  the asymmetry-based tuning of the valleytronic properties}.
\newblock {\em Phys. Rev. B}, 101:245416, Jun 2020.

\bibitem{kresse1996}
Georg Kresse and J{\"u}rgen Furthm{\"u}ller.
\newblock {Efficiency of ab-initio total energy calculations for metals and
  semiconductors using a plane-wave basis set}.
\newblock {\em Comput. Mater. Sci.}, 6:15--50, 1996.

\bibitem{Georg1996}
Georg Kresse and J{\"u}rgen Furthm{\"u}ller.
\newblock Efficient iterative schemes for ab initio total-energy calculations
  using a plane-wave basis set.
\newblock {\em Phys. Rev. B}, 54:11169--11204, 1996.

\bibitem{hohenberg1964}
Pierre Hohenberg and Walter Kohn.
\newblock Inhomogeneous electron gas.
\newblock {\em Phys. Rev.}, 136--143:B864, 1964.

\bibitem{kohn1965}
Walter Kohn and Lu~Jeu Sham.
\newblock Self-consistent equations including exchange and correlation effects.
\newblock {\em Phys. Rev.}, 140:A1133--A1138, 1965.

\bibitem{kresse1999}
Georg Kresse and Daniel Joubert.
\newblock From ultrasoft pseudopotentials to the projector augmented-wave
  method.
\newblock {\em Phys. Rev. B}, 59:1758--1775, 1999.

\bibitem{blochl1994}
Peter~E Bl{\"o}chl.
\newblock Projector augmented-wave method.
\newblock {\em Phys. Rev. B}, 50:17953--17979, 1994.

\bibitem{grimme2010}
Stefan Grimme, Jens Antony, Stephan Ehrlich, and Helge Krieg.
\newblock A consistent and accurate ab initio parametrization of density
  functional dispersion correction (\textsc{DFT-D}) for the 94 elements
  \textsc{H}-\textsc{P}u.
\newblock {\em J. Chem. Phys.}, 132:154104--154123, 2010.

\bibitem{monkhorst1976}
Hendrik~J Monkhorst and James~D Pack.
\newblock {Special points for Brillouin-zone integrations}.
\newblock {\em Phys. Rev. B}, 13:5188--5192, 1976.

\bibitem{perdew1996}
John~P Perdew, Kieron Burke, and Matthias Ernzerhof.
\newblock Generalized gradient approximation made simple.
\newblock {\em Phys. Rev. Lett.}, 77:3865--3868, 1996.

\bibitem{Liechtenstein1995}
A.~I. Liechtenstein, V.~I. Anisimov, and J.~Zaanen.
\newblock {Density-functional theory and strong interactions: Orbital ordering
  in Mott-Hubbard insulators}.
\newblock {\em Phys. Rev. B}, 52:R5467--R5470, Aug 1995.

\bibitem{Li2014}
Fengyu Li, Kaixiong Tu, and Zhongfang Chen.
\newblock {Versatile Electronic Properties of $VSe_{2}$ Bulk, Few-Layers,
  Monolayer, Nanoribbons, and Nanotubes: A Computational Exploration}.
\newblock {\em The Journal of Physical Chemistry C}, 118(36):21264--21274, Sep
  2014.

\bibitem{aryasetiawan2006}
F.~Aryasetiawan, K.~Karlsson, O.~Jepsen, and U.~Sch\"onberger.
\newblock {Calculations of Hubbard $U$ from first-principles}.
\newblock {\em Phys. Rev. B}, 74:125106, Sep 2006.

\bibitem{tang2009}
W~Tang, E~Sanville, and G~Henkelman.
\newblock {A grid-based \textsc{B}ader analysis algorithm without lattice
  bias}.
\newblock {\em J. Phys. Condens. Matter}, 21:084204, 2009.

\bibitem{AIMD1994}
G.~Kresse and J.~Hafner.
\newblock Ab initio molecular-dynamics simulation of the
  liquid-metal--amorphous-semiconductor transition in germanium.
\newblock {\em Phys. Rev. B}, 49:14251--14269, May 1994.

\bibitem{Andrew2012}
R.~C. Andrew, R.~E. Mapasha, A.~M. Ukpong, and N.~Chetty.
\newblock Mechanical properties of graphene and boronitrene.
\newblock {\em Phys. Rev. B}, 85:125428, Mar 2012.

\bibitem{WANG2021}
Vei Wang, Nan Xu, Jin-Cheng Liu, Gang Tang, and Wen-Tong Geng.
\newblock {VASPKIT: A user-friendly interface facilitating high-throughput
  computing and analysis using VASP code}.
\newblock {\em Computer Physics Communications}, 267:108033, 2021.

\bibitem{supporting-data}
{See Supplemental Material at [] for the detailed structural, electronic, and
  magnetic properties of Janus VSSe monolayer. For the VSSe bilayer, different
  magnetic configurations, its 3D band structure, charge density difference,
  and total density of states plots are presented.}

\bibitem{Qi2020}
Shengmei Qi, Jiawei Jiang, and Wenbo Mi.
\newblock {Tunable valley polarization{,} magnetic anisotropy and
  Dzyaloshinskii--Moriya interaction in two-dimensional intrinsic ferromagnetic
  Janus 2H-VSeX (X = S{,} Te) monolayers}.
\newblock {\em Phys. Chem. Chem. Phys.}, 22:23597--23608, 2020.

\bibitem{Wasey2015}
A.~H. M.~Abdul Wasey, Soubhik Chakrabarty, and G.~P. Das.
\newblock {Quantum size effects in layered $VX_{2}$ (X = S, Se) materials:
  Manifestation of metal to semimetal or semiconductor transition}.
\newblock {\em Journal of Applied Physics}, 117(6):064313, 02 2015.

\bibitem{wang2018}
Jun Wang, Haibo Shu, Tianfeng Zhao, Pei Liang, Ning Wang, Dan Cao, and
  Xiaoshuang Chen.
\newblock {Intriguing electronic and optical properties of two-dimensional
  Janus transition metal dichalcogenides}.
\newblock {\em Phys. Chem. Chem. Phys.}, 20:18571--18578, 2018.

\bibitem{He2014}
Jiangang He, Kerstin Hummer, and Cesare Franchini.
\newblock {Stacking effects on the electronic and optical properties of bilayer
  transition metal dichalcogenides ${\mathrm{MoS}}_{2}$, ${\mathrm{MoSe}}_{2}$,
  ${\mathrm{WS}}_{2}$, and ${\mathrm{WSe}}_{2}$}.
\newblock {\em Phys. Rev. B}, 89:075409, Feb 2014.

\bibitem{Nose1984}
Shuichi Nos{\'e}.
\newblock A unified formulation of the constant temperature molecular dynamics
  methods.
\newblock {\em The Journal of Chemical Physics}, 81(1):511--519, 07 1984.

\bibitem{Born1955}
Max Born, Kun Huang, and M.~Lax.
\newblock {Dynamical Theory of Crystal Lattices}.
\newblock {\em American Journal of Physics}, 23(7):474--474, 10 1955.

\bibitem{zeng2015}
Fan Zeng, Wei-Bing Zhang, and Bi-Yu Tang.
\newblock {Electronic structures and elastic properties of monolayer and
  bilayer transition metal dichalcogenides $MX_{2}$ (M= Mo, W; X= O, S, Se,
  Te): a comparative first-principles study}.
\newblock {\em Chinese Physics B}, 24(9):097103, 2015.

\bibitem{THANH2020}
Vuong~Van Thanh, Nguyen~Duy Van, Do~Van Truong, Riichiro Saito, and Nguyen~Tuan
  Hung.
\newblock {First-principles study of mechanical, electronic and optical
  properties of Janus structure in transition metal dichalcogenides}.
\newblock {\em Applied Surface Science}, 526:146730, 2020.

\bibitem{Lu2017}
Ang-Yu Lu, Hanyu Zhu, Jun Xiao, Chih-Piao Chuu, Yimo Han, Ming-Hui Chiu,
  Chia-Chin Cheng, Chih-Wen Yang, Kung-Hwa Wei, Yiming Yang, Yuan Wang,
  Dimosthenis Sokaras, Dennis Nordlund, Peidong Yang, David~A. Muller, Mei-Yin
  Chou, Xiang Zhang, and Lain-Jong Li.
\newblock Janus monolayers of transition metal dichalcogenides.
\newblock {\em Nature Nanotechnology}, 12(8):744--749, Aug 2017.

\bibitem{Chen2018}
P.~Chen, Woei~Wu Pai, Y.-H. Chan, V.~Madhavan, M.~Y. Chou, S.-K. Mo, A.-V.
  Fedorov, and T.-C. Chiang.
\newblock {Unique Gap Structure and Symmetry of the Charge Density Wave in
  Single-Layer ${\mathrm{VSe}}_{2}$}.
\newblock {\em Phys. Rev. Lett.}, 121:196402, Nov 2018.

\bibitem{vanEfferen2024}
Camiel van Efferen, Joshua Hall, Nicolae Atodiresei, Virginia Boix, Affan
  Safeer, Tobias Wekking, Nikolay~A. Vinogradov, Alexei~B. Preobrajenski, Jan
  Knudsen, Jeison Fischer, Wouter Jolie, and Thomas Michely.
\newblock {2D Vanadium Sulfides: Synthesis, Atomic Structure Engineering, and
  Charge Density Waves}.
\newblock {\em ACS Nano}, 18(22):14161--14175, Jun 2024.

\bibitem{goodenough2008}
John~B Goodenough.
\newblock {Goodenough-Kanamori rule}.
\newblock {\em Scholarpedia}, 3(10):7382, 2008.

\bibitem{Zha2022}
Jiajia Zha, Mingcheng Luo, Ming Ye, Tanveer Ahmed, Xuechao Yu, Der-Hsien Lien,
  Qiyuan He, Dangyuan Lei, Johnny~C. Ho, James Bullock, Kenneth~B. Crozier, and
  Chaoliang Tan.
\newblock {Infrared Photodetectors Based on 2D Materials and Nanophotonics}.
\newblock {\em Advanced Functional Materials}, 32(15):2111970, 2022.

\bibitem{Wang2008}
X.~L. Wang.
\newblock {Proposal for a New Class of Materials: Spin Gapless Semiconductors}.
\newblock {\em Phys. Rev. Lett.}, 100:156404, Apr 2008.

\bibitem{Yue2020}
Zengji Yue, Zhi Li, Lina Sang, and Xiaolin Wang.
\newblock {Spin-Gapless Semiconductors}.
\newblock {\em Small}, 16(31):1905155, 2020.

\bibitem{Weintrub2022}
Benjamin~I Weintrub, Yu-Ling Hsieh, Sviatoslav Kovalchuk, Jan~N Kirchhof,
  Kyrylo Greben, and Kirill~I Bolotin.
\newblock {Generating intense electric fields in {2D} materials by dual ionic
  gating}.
\newblock {\em Nat Commun}, 13(1):6601, November 2022.

\end{thebibliography}


\begin{thebibliography}{1}

\bibitem{Zhang2019}
Chunmei Zhang, Yihan Nie, Stefano Sanvito, and Aijun Du.
\newblock {First-Principles Prediction of a Room-Temperature Ferromagnetic
  Janus VSSe Monolayer with Piezoelectricity, Ferroelasticity, and Large Valley
  Polarization}.
\newblock {\em Nano Letters}, 19(2):1366--1370, Feb 2019.

\bibitem{Dey2020}
Dibyendu Dey and Antia~S. Botana.
\newblock {Structural, electronic, and magnetic properties of vanadium-based
  Janus dichalcogenide monolayers: A first-principles study}.
\newblock {\em Phys. Rev. Mater.}, 4:074002, Jul 2020.

\bibitem{Qi2020}
Shengmei Qi, Jiawei Jiang, and Wenbo Mi.
\newblock {Tunable valley polarization{,} magnetic anisotropy and
  Dzyaloshinskii--Moriya interaction in two-dimensional intrinsic ferromagnetic
  Janus 2H-VSeX (X = S{,} Te) monolayers}.
\newblock {\em Phys. Chem. Chem. Phys.}, 22:23597--23608, 2020.

\bibitem{Zhuang2016}
Houlong~L. Zhuang and Richard~G. Hennig.
\newblock {Stability and magnetism of strongly correlated single-layer
  $VS_{2}$}.
\newblock {\em Phys. Rev. B}, 93:054429, Feb 2016.

\bibitem{Wasey2015}
A.~H. M.~Abdul Wasey, Soubhik Chakrabarty, and G.~P. Das.
\newblock {Quantum size effects in layered $VX_{2}$ (X = S, Se) materials:
  Manifestation of metal to semimetal or semiconductor transition}.
\newblock {\em Journal of Applied Physics}, 117(6):064313, 02 2015.

\bibitem{Luo2020}
Chaobo Luo, Xiangyang Peng, Jinfeng Qu, and Jianxin Zhong.
\newblock {Valley degree of freedom in ferromagnetic Janus monolayer H-VSSe and
  the asymmetry-based tuning of the valleytronic properties}.
\newblock {\em Phys. Rev. B}, 101:245416, Jun 2020.

\bibitem{Zhang2017}
Jing Zhang, Shuai Jia, Iskandar Kholmanov, Liang Dong, Dequan Er, Weibing Chen,
  Hua Guo, Zehua Jin, Vivek~B. Shenoy, Li~Shi, and Jun Lou.
\newblock {Janus Monolayer Transition-Metal Dichalcogenides}.
\newblock {\em ACS Nano}, 11(8):8192--8198, Aug 2017.

\bibitem{wang2018}
Jun Wang, Haibo Shu, Tianfeng Zhao, Pei Liang, Ning Wang, Dan Cao, and
  Xiaoshuang Chen.
\newblock Intriguing electronic and optical properties of two-dimensional janus
  transition metal dichalcogenides.
\newblock {\em Phys. Chem. Chem. Phys.}, 20:18571--18578, 2018.

\end{thebibliography}

\end{document}


\title{Supplemental Material for Electric field induced half-metallicity
in a two-dimensional ferromagnetic Janus VSSe bilayer}
\author{Khushboo Dange}
\email{khushboodange@gmail.com}

\affiliation{Department of Physics, Indian Institute of Technology Bombay, Powai,
Mumbai 400076, India}
\author{Shivprasad S. Shastri}
\email{shastri1992@gmail.com}

\affiliation{Department of Physics, Indian Institute of Technology Bombay, Powai,
Mumbai 400076, India}
\author{Alok Shukla}
\email{shukla@iitb.ac.in}

\affiliation{Department of Physics, Indian Institute of Technology Bombay, Powai,
Mumbai 400076, India}
\maketitle

\subsection{Ferromagnetic (FM) Janus 2H-VSSe Monolayer}

\subsubsection{Structural Properties}

\begin{figure}[H]
\begin{centering}
\includegraphics[scale=0.4]{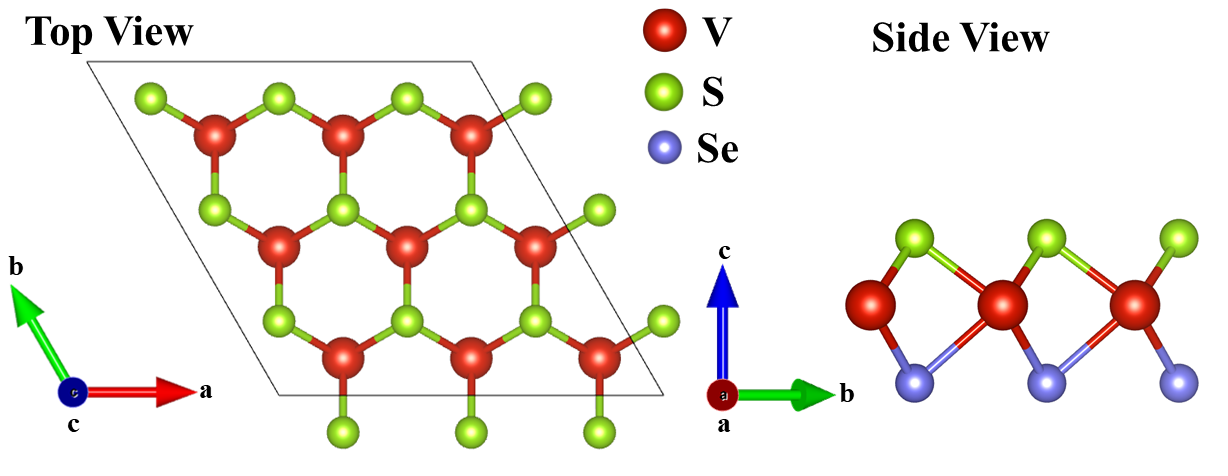} 
\par\end{centering}
\caption{Crystal structure of the 2H polymorph of the Janus VSSe monolayer.}
\label{fig:mono_struc} 
\end{figure}

The Janus VSSe monolayer has been theoretically predicted to exhibit
intrinsic 2D ferromagnetism, with FM order primarily contributed by
the V-3\textit{d} electrons \cite{Zhang2019,Dey2020}. Similar to
non-Janus TMDs, two structural forms of the Janus VSSe monolayer have
been predicted: (a) trigonal prismatic (2H) and (b) octahedral (1T),
with the 2H-phase being energetically most stable \cite{Dey2020}.
Both the 2H and 1T phases of VSSe monolayer have \textit{P3m1} space
group. The crystal structure of the 2H polymorph of the VSSe monolayer
possessing $C_{3v}$ symmetry is depicted in Fig. \ref{fig:mono_struc}.
The Wyckoff position of vanadium atom is 1\textit{a} (0, 0, z) and
of chalcogens is 1\textit{c} (2/3, 1/3, z). The optimized in-plane
lattice constant of the considered 2H-VSSe monolayer is 3.25 {\AA},
consistent with the reported ones \cite{Qi2020}, and lies between
the lattice constant of VS$_{2}$ (3.12 {\AA}) and VSe$_{2}$ (3.33
{\AA}) monolayers \cite{Zhuang2016,Wasey2015}. The V-S and V-Se
bond lengths are 2.37 {\AA} and 2.51 {\AA}, respectively, consistent
with the reported values \cite{Qi2020}, and are in agreement with
the size of the anions i.e., $r_{S}<r_{Se}$ ($r$ denotes the ionic
radius). To predict the energetic stability of the considered Janus
VSSe monolayer, we have estimated its formation energy ($E_{f}^{mono}$)
using Eq. \ref{eq:form-mono} by assuming that 2D Janus VSSe can be
derived from VS\textsubscript{2} and VSe\textsubscript{2}. 
\begin{equation}
E_{f}^{mono}=E_{VSSe}-\frac{1}{2}(E_{VS_{2}}+E_{VSe_{2}})\label{eq:form-mono}
\end{equation}

where $E_{VS_{2}}$, $E_{VSe_{2}}$, and $E_{VSSe}$ are the energies
of the optimized VS\textsubscript{2}, VSe\textsubscript{2}, and
VSSe monolayers. These energies and the resultant $E_{f}^{mono}$
are reported in Table \ref{tab:energies-mono} in the units of eV
per formula unit (eV/f.u.). Our calculated negative formation energy
(-0.046 eV/f.u.) suggests that the Janus VSSe monolayer can be formed
in equilibrium environment. We note that Luo \emph{et al}. \cite{Luo2020}
also reported a negative formation energy of -0.00263 eV/f.u., which
is about an order of magnitude smaller than our value. For comparison,
we consider the case of Janus MoSSe monolayer which has been synthesized
experimentally \cite{Zhang2017}, although its formation energy has
been computed to have a positive value of 0.02699 eV by Luo \emph{et
al}. \cite{Luo2020}. Therefore, according to our calculations, VSSe
monolayer can also be synthesized, because it is predicted to be more
stable as compared to MoSSe. The cohesive energy per atom, $E_{c}^{mono}$
is also calculated using the formula

\begin{equation}
E_{c}^{mono}=\frac{1}{3}\left(E_{VSSe}-(E_{V}+E_{S}+E_{Se})\right)\label{eq:form-mono-1}
\end{equation}

where $E_{V}$, $E_{S}$, and $E_{Se}$ represent the energies of
the isolated V, S, and Se atoms, respectively. The resultant $E_{c}^{mono}$
of -4.56 eV is about twice of that reported for the MoSSe monolayer
(-2.32 eV) \cite{wang2018}, and thus hints at the thermodynamic stability
of the 2D Janus VSSe structure. Furthermore, its dynamic stability
has already been confirmed in various studies through ab initio molecular
dynamics (AIMD) simulations \cite{Luo2020}, as well as by the absence
of imaginary frequencies in the phonon dispersion \cite{Zhang2019,Luo2020,Dey2020}.

\begin{table}[H]
\caption{Total energies (eV/f.u.) of the involved systems, formation energy,
$E_{f}^{mono}$ (eV/f.u.), and cohesive energy per atom, $E_{c}^{mono}$
(eV) of the monolayer VSSe.}
\label{tab:energies-mono}

\centering{}%
\begin{tabular}{ccccc}
\toprule 
$E_{VS_{2}}$  & $E_{VSe_{2}}$  & $E_{VSSe}$  & $E_{f}^{mono}$  & $E_{c}^{mono}$ \tabularnewline
\midrule 
-19.703  & -17.990  & -18.892  & -0.046  & -4.560\tabularnewline
\bottomrule
\end{tabular}
\end{table}

\subsubsection{Magnetic Moments}

We computed the magnetic moments associated with each atomic species
using the GGA+\textit{U} method. The resultant magnetic moments associated
with S, Se, and V atoms are -0.08$\mu_{B}$, -0.14$\mu_{B}$, and
1.19$\mu_{B}$, respectively, which are consistent with the reported
values \cite{Dey2020}.

\subsubsection{Electronic Properties}

\begin{figure}[H]
\begin{centering}
\includegraphics[scale=0.6]{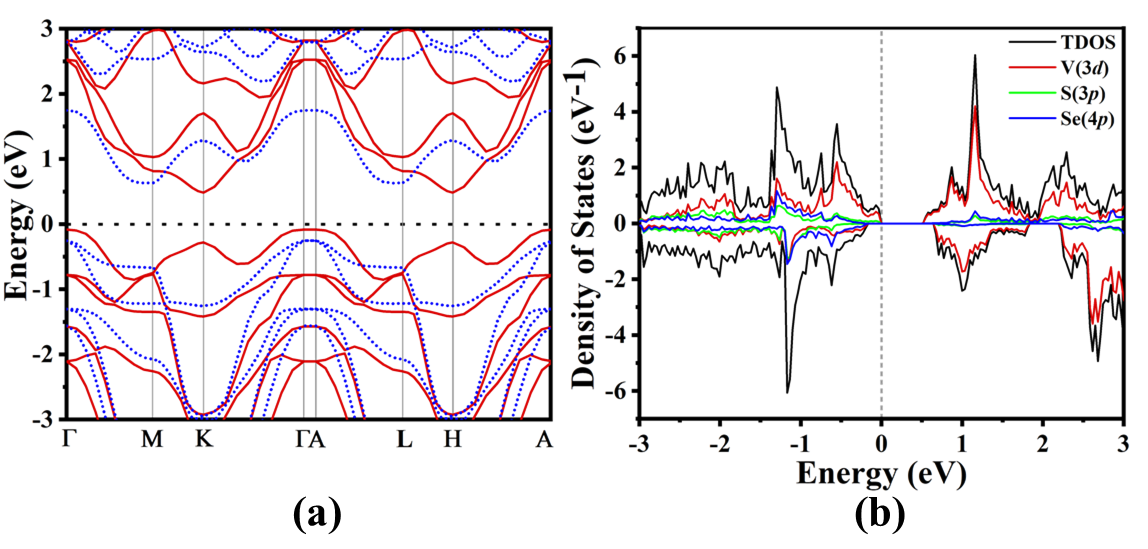} 
\par\end{centering}
\caption{(a) Electronic band structure and (b) density of states (DOS) computed
for the FM Janus VSSe monolayer. The solid red and dotted blue lines
in (a) represent the energy states corresponding to the majority and
minority charge carriers, respectively.}
\label{fig:mono-bands} 
\end{figure}

\begin{figure}[H]
\begin{centering}
\includegraphics[scale=0.4]{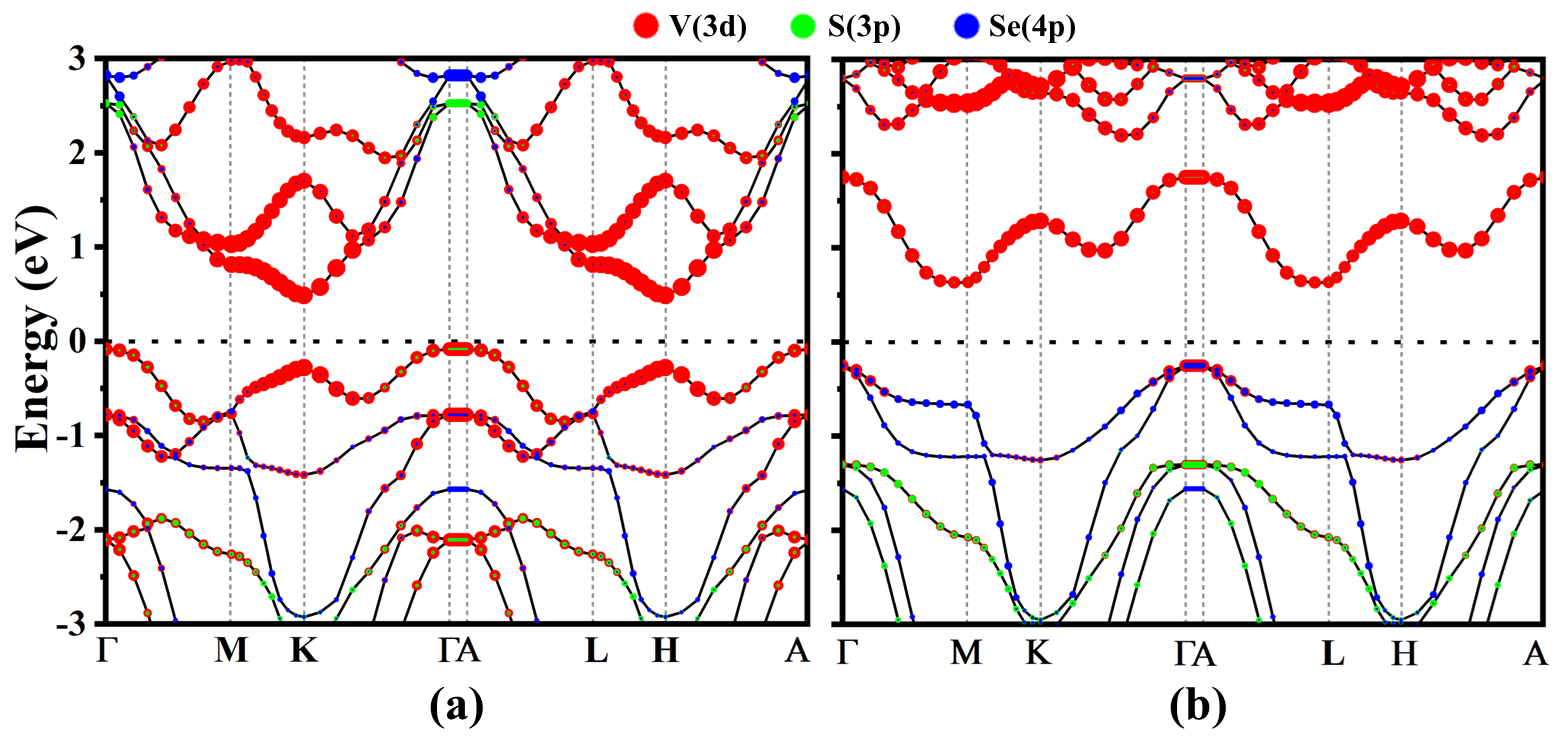} 
\par\end{centering}
\caption{Projected band structures for (a) majority and (b) minority spin carriers
of the Janus VSSe monolayer.}
\label{fig:mono_pband} 
\end{figure}

The electronic band structure simulations are performed within the
spin-polarized formalism for the considered FM Janus VSSe monolayer
using the GGA+\textit{U} method, and the results are presented in
Fig. \ref{fig:mono-bands}(a). The band gaps ($E_{g}$) of 0.57 eV
and 0.88 eV are obtained for the majority and minority spin carriers,
respectively. It is to be noted that the valence band maximum (VBM)
for both the spin carriers is positioned at the same high symmetry
point, i.e., \textit{$\Gamma$}, whereas, the conduction band minima
(CBM) occur at different \textbf{k}-points. For the majority (minority)
spins, CBM is positioned at \textit{K} (close to \textit{L} along
\textit{L}$\rightarrow$\textit{A}), and thus an indirect $E_{g}$
results for both the spin carriers. Our calculated band structure
agrees well with that reported in the literature \cite{Dey2020}.
The total density of states (TDOS) and atom-projected partial density
of states (PDOS) decomposed by azimuthal quantum number (\textit{l})
are also computed, as depicted in Fig. \ref{fig:mono-bands}(b). The
asymmetric nature of the DOS can be observed for the FM ground state
of the Janus VSSe monolayer. The PDOS suggests that the CBM of both
spin carriers comprises mainly of V-3\textit{d} orbitals. The VBM
corresponding to majority spin carriers is mainly composed of V-3\textit{d}
orbitals, whereas, the hybridization of the Se-4\textit{p} and V-3\textit{d}
orbitals is involved in the VBM of minority states. The atom-projected
\textit{l}-decomposed band structures are also presented in Fig. \ref{fig:mono_pband}.
This plot supports the orbital characters of the VBM and CBM as shown
by the PDOS plot.

\begin{figure}[H]
\begin{centering}
\includegraphics[scale=0.4]{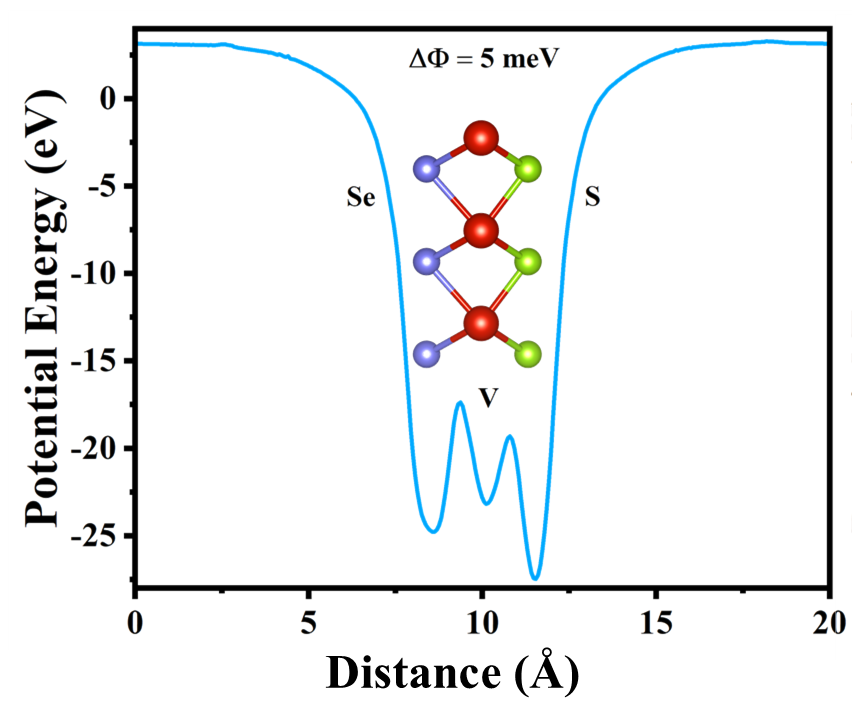} 
\par\end{centering}
\caption{Planar average of the electrostatic potential energy as a function
of distance along z-direction.}
\label{fig:potential_mono} 
\end{figure}

The broken horizontal mirror symmetry within the Janus VSSe monolayer
gives rise to spontaneous polarization normal to the plane, which
is attributed to the inherent electric field (\textbf{$\boldsymbol{E_{in}}$}).
From the Bader charge analysis, we found that S and Se atoms gain
0.70e and 0.54e charges, respectively, from the V atom which loses
1.24e charge. It is clear that due to the electronegativity order:
S (2.58) > Se (2.55) > V(1.63), S atom gains more electronic charge
from V atom as compared to the Se atom, resulting in \textbf{$\boldsymbol{E_{in}}$}
along the positive z-direction (upwards, i.e., from Se to S). The
planar average of the electrostatic potential energy is taken along
the $z$-direction as shown in Fig. \ref{fig:potential_mono}. From
Fig. \ref{fig:potential_mono}, we can calculate the change in work
function $\Delta\Phi$ as we go from Se to S, and its value turns
out to be nonzero (5 meV), further confirming the presence of $\boldsymbol{E_{in}}$.

\subsection{Janus 2H-VSSe Bilayer}

\subsubsection{Different magnetic configurations and stability}

\begin{figure}
\begin{centering}
\includegraphics[scale=0.65]{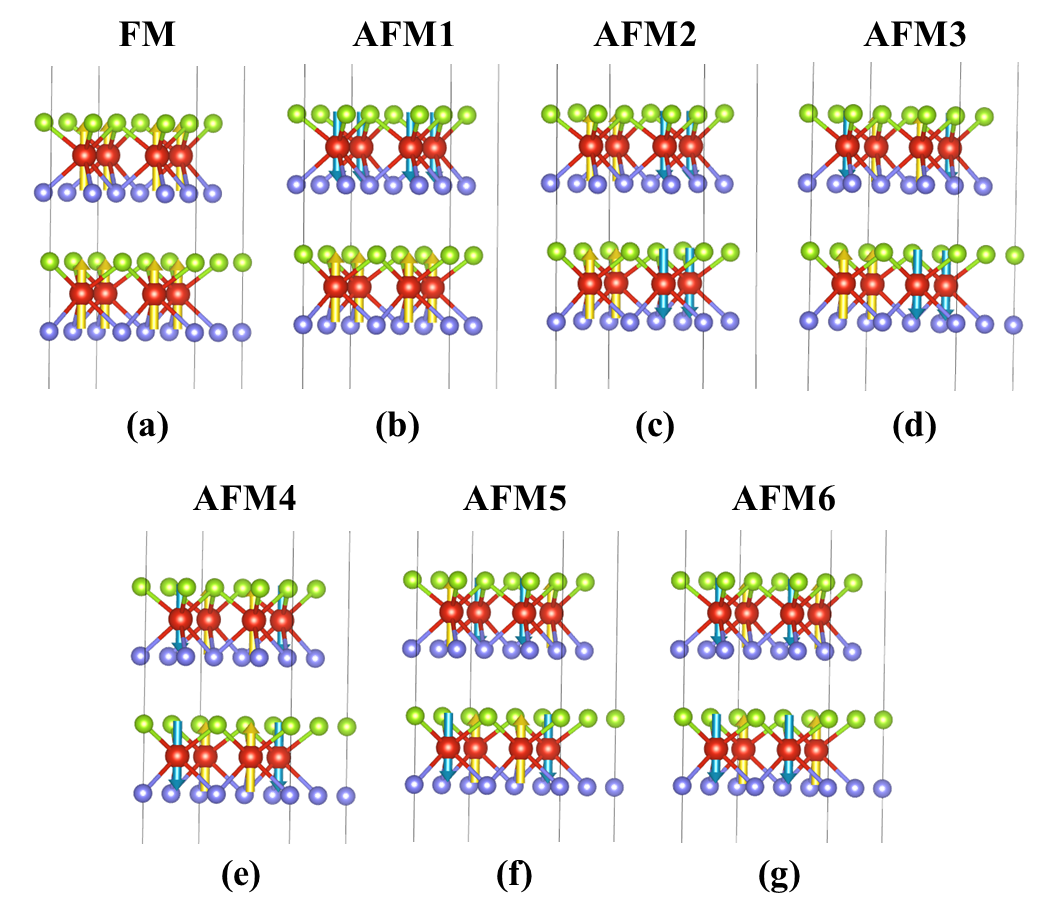} 
\par\end{centering}
\caption{Different magnetic configurations considered for the AB stacked bilayer.
Yellow and blue vertical arrows denote the majority and minority spins
localized on Vanadium atoms, respectively. The same magnetic configurations
were used for all the stacking arrangements considered in this work.}
\label{mag-confg} 
\end{figure}

Using 2x2 supercells of each of the considered stacking type, we have
constructed seven different magnetic configurations which include
an FM and six antiferromagnetic (AFM) configurations (see Fig. \ref{mag-confg}).
The FM configuration possesses both interlayer and intralayer ferromagnetic
coupling, with all the spins aligned in the same direction (Fig. \ref{mag-confg}(a)).
The AFM1 configuration (Fig. \ref{mag-confg}(b)) is designed such
that it is intralayer FM (spins within a layer are aligned in the
same direction) but interlayer AFM (spins of the two layers are equal
and aligned opposite to each other). In the other considered AFM configurations,
AFM2 - AFM6 (Fig. \ref{mag-confg}), the spins within each single
layer are aligned opposite to each other and thus possess both inter-
and intralayer AFM characteristics. 

\subsubsection{3D Band Structure}

\begin{figure}[H]
\begin{centering}
\includegraphics{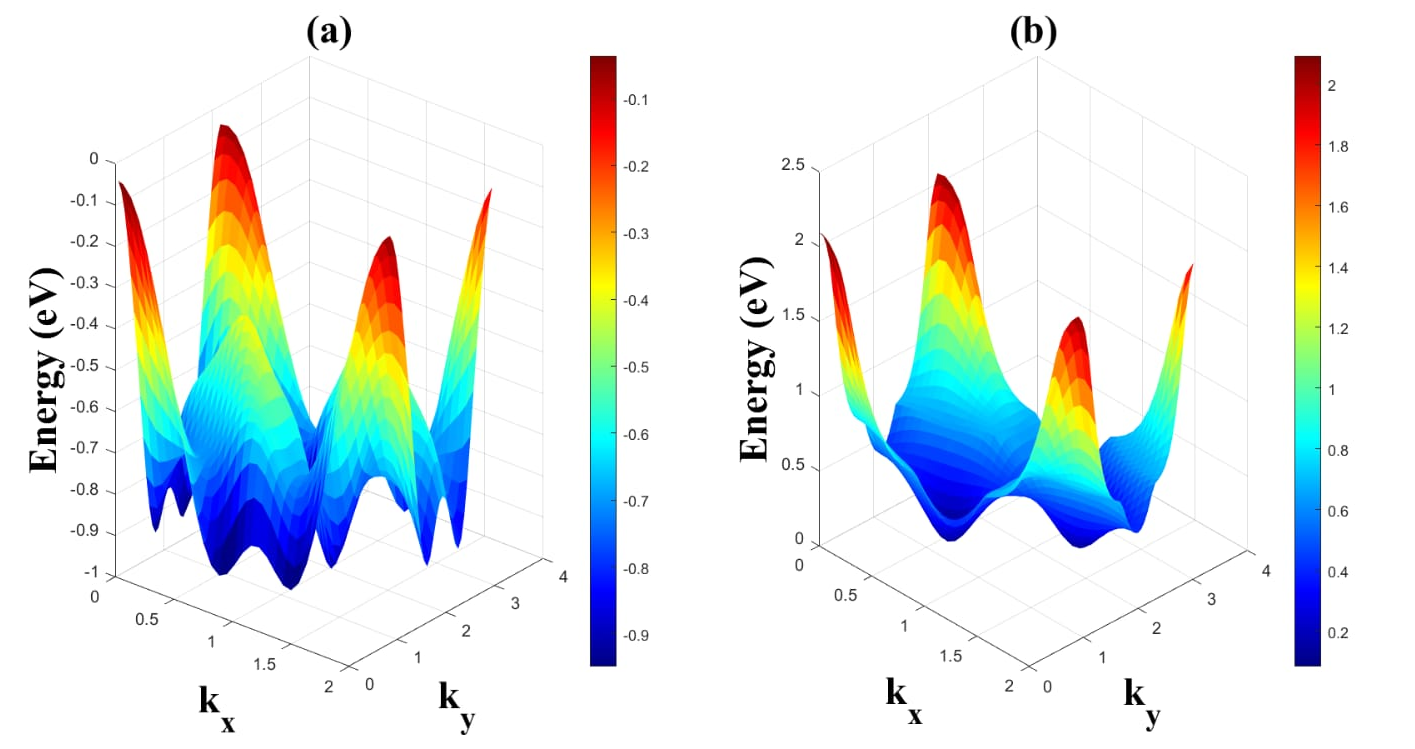}
\par\end{centering}
\caption{3D plots of the (a) highest valence band and (b) lowest conduction
band corresponding to the majority spin carriers at $k_{z}$ = 0 plane
with $E_{F}$ shifted to zero.}
\end{figure}

\subsubsection{Effect of external electric field on the electronic properties}

\begin{figure}[H]
\begin{centering}
\includegraphics[scale=0.5]{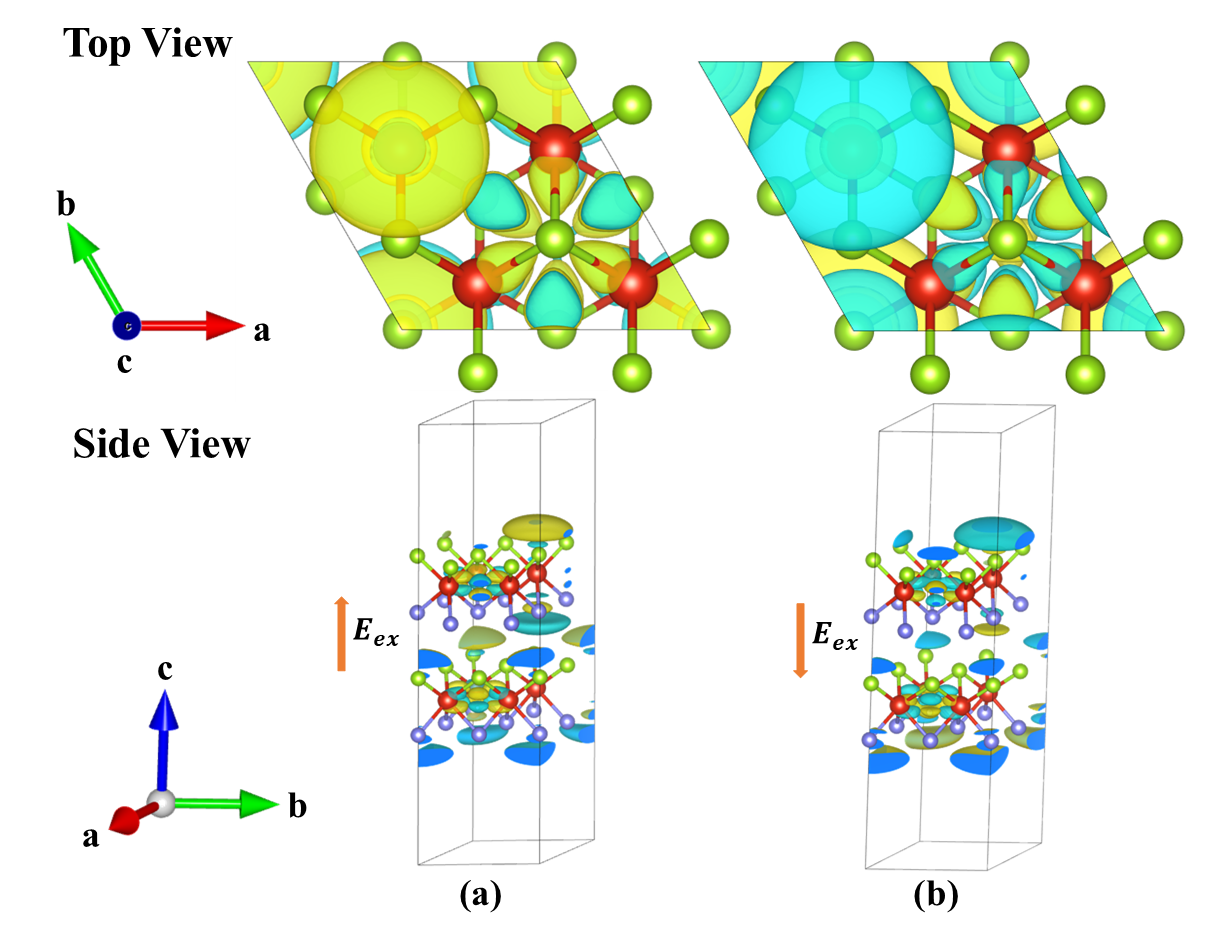} 
\par\end{centering}
\caption{Charge density difference between the cases $\boldsymbol{E_{ex}}=0.0$
and $\boldsymbol{E_{ex}}=0.18$ V/{\AA} applied in the (a) upward
and (b) downward directions of the considered VSSe bilayer. The yellow
and blue colors correspond to the positive and negative isosurfaces
at $1.27\times10^{-4}$ e/{\AA$^{3}$}, and depict the charge accumulation
and depletion regions, respectively.}
\end{figure}

\begin{figure}[H]
\begin{centering}
\includegraphics[scale=0.5]{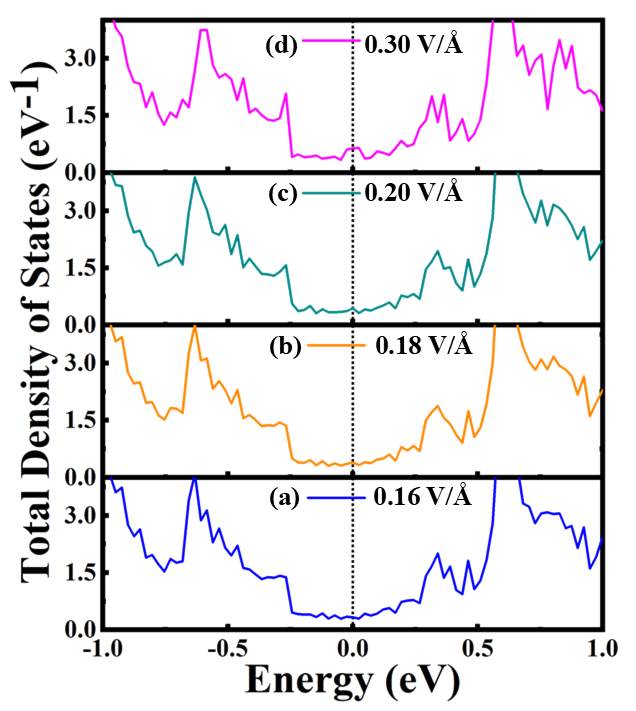} 
\par\end{centering}
\caption{Total density of states (TDOS) of the majority spin carriers of the
VSSe bilayer in the presence of external electric fields of intensities:
(a) 0.16 V/{\AA}, (b) 0.18 V/{\AA}, (c) 0.20 V/{\AA}, and
(d) 0.30 V/{\AA}, applied downward.}
\end{figure}

\bibliographystyle{unsrt}
\bibliography{Ref-SI}